\documentclass[journal]{IEEEtran}

\usepackage[T1]{fontenc} 

\usepackage{amsmath,amsfonts,amssymb}
\usepackage{mathtools}
\usepackage{mathrsfs}
\usepackage{bm} 
\usepackage{cases}

\usepackage{graphicx}
\usepackage{booktabs} 
\usepackage{multicol}
\usepackage{multirow}
\usepackage{diagbox}
\usepackage{makecell}
\usepackage{threeparttable} 
\usepackage[caption=false,font=footnotesize,labelfont=sf,textfont=rm]{subfig} 
\usepackage{stfloats} 
\usepackage{setspace} 
\usepackage{indentfirst}
\usepackage{balance} 
\usepackage[normalem]{ulem} 
\usepackage{color}

\usepackage{algorithm}
\usepackage{algorithmic}

\usepackage{cite} 
\usepackage{url}
\usepackage{textcomp}
\usepackage{verbatim}
\usepackage{enumerate}
\usepackage{utfsym}
\usepackage{tikz}
\usepackage{array}
\usepackage{orcidlink} 
\usepackage[doipre={doi:~}]{uri}

\hypersetup{
	colorlinks=true,
	linkcolor=black,
	citecolor=black,
	urlcolor=black
}

\newtheorem{theorem}{\bf{Theorem}}
\newtheorem{remark}{\bf{Remark}}
\newtheorem{lemma}{\bf{Lemma}}

\newtheorem{property}{\bf{Property}}
\newtheorem{definition}{\bf{Definition}}

\def\BibTeX{{   B\kern-.05em{\sc i\kern-.025em b}\kern-.08em
		T\kern-.1667em\lower.7ex\hbox{E}\kern-.125emX}}

\begin{document}
	
	\title{QPSAN: Quantum Parameterized Self-Attention Network for Image Classification}
\author{Wenwei Zhang$^{\orcidlink{0009-0003-5734-7670}}$, Jintao Wang$^{\orcidlink{0009-0003-4712-5353}}$, Tianyu Ye$^{\orcidlink{0000-0002-5581-9895}}$ and Changgeng Liao$^{\orcidlink{0000-0003-0104-3814}}$ 
	\thanks{
		This work was supported in part by the National Natural Science Foundation of China (Grants No. 12004336, No. 12075205,and No. 62071430); in part by the Fundamental Research Funds for the Provincial University of Zhejiang (Grant No. XRK23006); and in part by Funds from the China Scholarship Council. (Corresponding author: Chang-Geng Liao; Tianyu Ye.)
		
		Wenwei Zhang, Jintao Wang and Tianyu Ye are with the School of Information and Electronic Engineering (Sussex Artificial Intelligence Institute), Zhejiang Gongshang University, Hangzhou, Zhejiang 310018, China (e-mail: 24020090040@pop.zjgsu.edu.cn; 1811080102@pop.zjgsu.edu.cn; yetianyu@zjgsu.edu.cn).
		
		Changgeng Liao is with the School of Information and Electronic Engineering (Sussex Artificial Intelligence Institute), Zhejiang Gongshang University, Hangzhou, Zhejiang 310018, China and also with Department of Physics and Astronomy, University of Sussex, Brighton BN1 9QH, United Kingdom (E-mail: cgliao@zjgsu.edu.cn).
		
		Our code is available at \href{https://github.com/Comets9224/QPSAN.git}{https://github.com/Comets9224/QPSAN.git}.
	}
}

\maketitle
	
\begin{abstract}
	Transformer now underpins modern AI as its core infrastructure.
	Its defining capability---dynamically focusing on the most relevant
	information in complex inputs---is bounded above by the self-attention
	scoring function. Quantum computing, with its superposition,
	entanglement, and probabilistic outputs, offers a fundamentally
	distinct computational framework for exploring beyond the design
	constraints of classical scoring functions. While quantum attention
	mechanisms have shown initial promise, existing works remain largely
	confined to redefining feature similarity measures, leaving the
	systematic use of parameterized quantum circuits (PQCs) as scoring
	functions largely unexplored; a substantial portion of existing schemes
	further rely on purely quantum architectures, precluding effective
	encoding of high-dimensional image inputs in the Noisy
	Intermediate-Scale Quantum era. We propose the Quantum
	Parameterized Self-Attention Network (QPSAN), implementing the
	self-attention scoring function via PQCs with only 5 trainable quantum
	parameters per layer. QPSAN computes query--key attention scores
	through quantum state encoding and joint measurement, yielding
	naturally bounded outputs without the explicit scaling of classical
	dot-product attention. We further establish a theoretical framework
	of the mathematical properties of this scoring function, demonstrating
	its potential to capture complex nonlinear query--key interactions, and
	quantifying the structural constraints of the encoding layer via
	effective degrees of freedom analysis. Experiments on four vision
	datasets show that QPSAN significantly outperforms the Vision
	Transformer (ViT) baseline, with the quantum representational advantage
	amplifying as data complexity increases. Ablation studies indicate that
	the performance gains may stem from the structural inductive bias of
	the quantum circuit rather than from parameter scale.
\end{abstract}
	
	
\section{Introduction}
\label{sec:introduction}

\IEEEPARstart{T}{he} Transformer architecture has driven breakthroughs 
in natural language processing~\cite{vaswani2017attention} and computer 
vision~\cite{dosovitskiy2020image}, among other domains, 
through its self-attention mechanism.
Self-attention measures inter-element relevance via query--key dot
products, weighting Value vectors to achieve context-aware dynamic
feature fusion. However, its output range is unbounded: attention scores 
drift with input magnitude, typically requiring a scaling factor $1/\sqrt{d_k}$ for correction, 
yet numerical instability may still arise under extreme inputs. 
Beyond this, the linear dot-product operation theoretically
constrains the capacity to model complex nonlinear interaction
patterns~\cite{henry2020query}.

Unlike classical mechanisms that rely on linear operations, quantum
computing constructs an attention paradigm grounded in state
superposition and entanglement. Li et al.~\cite{li2024quantum} pioneered
quantum self-attention for text classification, spurring a series of
quantum attention architectures. The Quantum Kernel Self-Attention
Network (QKSAN)~\cite{zhao2024qksan} introduces quantum kernel functions
into Hilbert space, replacing the classical dot product with kernel
similarity. The Quantum Self-Attention Network (QSAN)~\cite{shi2024qsan}
bypasses intermediate measurements, defining similarity directly on
quantum states via logical operations. The Hybrid Quantum Vision
Transformer (HQViT)~\cite{zhang2025hqvit} estimates quantum state
overlap via quantum swap tests, replacing the classical dot product from
a different angle. While these works have advanced quantum attention
modeling from different perspectives, they remain confined to redefining
feature similarity measures, providing no structural remedy for either
the unbounded range or the limited expressiveness of classical dot
products. Most existing schemes further rely on fully quantum
architectures, making effective encoding of high-dimensional image
inputs difficult under the hardware constraints of the Noisy
Intermediate-Scale Quantum (NISQ) era.

Parameterized quantum circuits (PQCs), with trainable rotation gates
and compatibility with hybrid architectures, can realize complex feature
interactions via shallow circuits---offering a viable path toward
embedding quantum modules directly into classical attention scoring
functions. Yet this direction remains systematically unexplored, lacking
both a theoretical characterization of its mathematical properties and a
structural-constraint analysis elucidating the source of its advantage
over classical alternatives.

To address these gaps, we propose the Quantum Parameterized
Self-Attention Network (QPSAN), for the first time directly employing
the joint measurement probability of PQCs as the attention scoring
function in place of the classical scaled dot product. We provide a
theoretical characterization of the circuit's mathematical properties
and elucidate the source of its difference from classical alternatives
through structural-constraint analysis. Our main contributions are as
follows:

\begin{enumerate}
	\item We design a two-qubit entangling circuit as the Quantum
	Parametric Attention (QPA) module, using the joint measurement
	probability $P(|00\rangle)+P(|11\rangle)$ as per-dimension attention
	scores. The output is naturally bounded in $[0,1]$, requiring no
	additional scaling, and QPSAN introduces only 5 trainable quantum
	parameters per Transformer block.
	
	\item We systematically analyze the mathematical properties of QPA
	as a scoring function---including boundedness, asymmetry, and
	non-monotonicity---revealing its potential to model nonlinear
	query--key interactions; we further investigate the structural
	inductive bias distinguishing QPA from classical dot-product
	attention and multilayer perceptron (MLP) scoring functions via
	effective degrees of freedom and non-separable kernel analysis.
	
	\item We conduct systematic comparative experiments on four benchmark
	datasets (FashionMNIST, DirtyMNIST, CIFAR-10, and FER2013),
	assessing statistical significance via paired $t$-tests over multiple
	runs, with QPSAN achieving $+3.54\%$ over the Vision Transformer
	(ViT) baseline on FER2013 ($p<0.001$); ablation experiments on MLP
	parameter efficiency further suggest that the contribution of the
	quantum circuit structure to performance gains cannot be replaced by
	classical parameter scaling.
\end{enumerate}

	The remainder of this paper is organized as follows.
	Section~\ref{sec:related_work} surveys prior work across three
	directions: attention mechanisms, variational quantum algorithms, and
	quantum attention. Section~\ref{sec:proposed_method} details the QPA
	circuit design, attention mechanism implementation, and theoretical
	analysis within the QPSAN architecture.
	Sections~\ref{sec:experiments} and~\ref{sec:results_and_analysis}
	present the experimental setup, results, and ablation studies on four
	datasets. Section~\ref{sec:conclusion} concludes the paper and outlines
	directions for future research.
	
	\section{Related Work}
	\label{sec:related_work}
	
	\subsection{Classical Attention}
	\label{subsec:classical_attention}
	
	Attention mechanisms, first introduced in natural language processing,
	enable models to dynamically focus on the most relevant portions of
	their inputs. In 2014, Bahdanau et al.~\cite{bahdanau2014neural}
	introduced attention into the encoder--decoder framework for neural
	machine translation, breaking the information bottleneck of compressing
	variable-length sentences into fixed-length vectors.
	Luong et al.~\cite{luongEffectiveApproachesAttentionbased2015} proposed
	global and local attention architectures,
	achieving then-state-of-the-art performance on English--German translation.
	Vaswani et al.~\cite{vaswani2017attention} proposed the Transformer,
	replacing recurrent and convolutional structures with multi-head
	self-attention; its high parallelism substantially accelerated training,
	cementing it as the shared backbone of modern large language models and
	vision foundation models. These three works introduced diverse scoring
	functions---additive, dot-product, bilinear, concatenation-based, and
	scaled dot-product---among which the scaled dot product has become
	dominant due to its computational efficiency.
	
	Attention mechanisms were subsequently extended to computer vision.
	In 2018, Non-local Neural Networks captured global responses by
	computing weighted sums of features across all spatial positions,
	enabling long-range spatial dependency modeling~\cite{wang2018non}.
	ViT, the baseline of this work, partitions images into patch sequences
	and feeds them into a pure Transformer, surpassing the then-best
	convolutional networks after large-scale pretraining~\cite{dosovitskiy2020image}.
	Subsequent works have driven the practical adoption of vision
	Transformers from different
	angles~\cite{liu2021swin,touvron2021training,pu2025linear,zheng2025linear}.
	
	Researchers then began systematically examining and refining attention
	mechanism design. Brauwers \& Frasincar~\cite{brauwers2021general}
	surveyed and categorized attention mechanisms across deep learning,
	providing a general analytical framework spanning multiple domains.
	Zhang et al.~\cite{zhangsurvey} focused further on efficient attention,
	organizing existing approaches into four paradigms---hardware-efficient,
	sparse, compact, and linear---and providing a unified mathematical
	framework for addressing the quadratic computation and memory
	bottlenecks of long-context models.
	
	Attention mechanism design remains an active research area; 
	yet all existing improvements operate within the classical computational 
	framework---that is, the basic form and mathematical properties of the 
	scoring function remain dominated by linear operations. This naturally raises
	a question: does an alternative computational paradigm exist that
	endows the scoring function with structural properties difficult to
	achieve directly within the classical framework?
	
	\subsection{Variational Quantum Algorithms}
	\label{subsec:variational_quantum_algorithm}
	
	\begin{figure}[!t]
		\centering
		\includegraphics[width=\columnwidth]{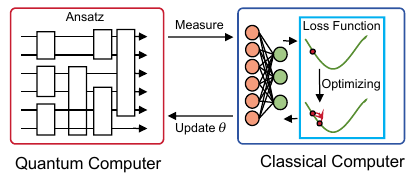}
		\caption{The \mbox{VQA} hybrid computational architecture: the quantum
			computer handles quantum state manipulation, while the classical
			computer performs parameter optimization and iterative updates.}
		\label{fig:vqa}
	\end{figure}
	
	Quantum computing offers a new direction for answering this question.
	As illustrated in Fig.~\ref{fig:vqa}, variational quantum algorithms
	(VQAs) constitute a hybrid computational paradigm coupling PQCs with
	classical optimizers: PQCs prepare and manipulate quantum states in the
	quantum state space, while classical optimizers iteratively update
	circuit parameters to minimize the cost function. Shallow circuit
	designs constrain circuit depth within hardware noise tolerance,
	enabling VQAs to produce meaningful results on NISQ devices. 
	On hardware feasibility, Kim et al.~\cite{kim2023evidence} demonstrated
	in 2023 on a 127-qubit processor that zero-noise extrapolation effectively
	extracts expectation values at circuit scales exceeding classical
	brute-force enumeration limits, with leading classical tensor-network
	approximations failing in strongly entangled regimes---providing critical
	empirical evidence for practical quantum computation in the
	pre-fault-tolerant era.
	
	VQAs offer three theoretical advantages. First, the exponential
	dimensionality of quantum state space may enhance model expressiveness
	at comparable parameter counts~\cite{abbas2021power}. Second,
	supervised quantum machine learning models are mathematically equivalent
	to kernel methods, in principle recasting training as a convex
	optimization problem and providing a path to circumvent barren
	plateaus~\cite{schuld2021supervised}. Third, the introduction of 
	quantum entanglement further yields provable
	quantum--classical learning advantages~\cite{zhaoEntanglementinducedProvableRobust2025}.
	
	On the application front, since Peruzzo et al.~\cite{peruzzo2014variational}
	introduced the Variational Quantum Eigensolver (VQE) in 2014, VQAs
	have been broadly applied to quantum chemistry, combinatorial
	optimization, and machine learning~\cite{cerezoVariationalQuantumAlgorithms2021}.
	In machine learning, quantum feature maps enable classification without
	quantum random access memory (QRAM), and quantum circuit learning
	circumvents the need for deep circuits via classical--quantum hybrid
	optimization~\cite{havlivcek2019supervised,mitarai2018quantum}. Recent
	work has further extended VQAs to robustness
	analysis~\cite{chenSuperiorResiliencePoisoning2026} and
	quantum-enhanced image
	classification~\cite{zhang2025dual}. At the circuit design level,
	systematic design criteria~\cite{sim2019expressibility} strengthen
	near-term feasibility, while the barren plateau phenomenon remains the
	central scalability challenge facing VQAs~\cite{larocca2025barren}.
	
	VQAs represent one of the most practically valuable paradigms in
	near-term quantum machine learning; building on this foundation,
	implementing attention-like mechanisms via PQCs has emerged as an
	active research direction.
	
	\subsection{Quantum Attention}
	\label{subsec:quantum_attention}
	
	Quantum attention mechanisms integrate quantum computing into attention
	modeling, aiming to leverage quantum superposition, entanglement, and
	parallelism to enhance the expressiveness of attention. Mirroring the
	trajectory of classical attention, quantum attention first emerged 
	in natural language processing (NLP) in 2024. Li et al.~\cite{li2024quantum}
	pioneered the Quantum Self-Attention Neural Networks model (QSANN),
	outperforming the then-best quantum NLP models on text classification.
	Subsequently, the Compact Learnable All-Quantum 
	Token Mixer with Shared-ansatz (CLAQS)~\cite{chen2025claqs}, the Quantum-inspired Text
	Sentiment Analysis model (QITSA)~\cite{li2024quantuma}, and
	Quantum-Inspired Self-Attention (QISA)~\cite{kuznetsov2026quantum}
	further refined self-attention from the perspectives of learnable
	quantum token mixing, complex-valued representations, and quantum state
	evolution, respectively. In the vision domain, Cherrat
	et al.~\cite{kerenidis2024quantum} achieved competitive performance on
	medical image classification. QKSAN~\cite{zhao2024qksan} combined
	quantum kernel methods with self-attention, demonstrating certain
	learning advantages with far fewer parameters than classical
	counterparts. QSAN~\cite{shi2024qsan} eliminated intermediate
	measurements via quantum logic similarity (QLS). The Quantum
	Complex-Valued Self-Attention Model (QCSAM)~\cite{chen2025quantuma}
	and the Quantum Mixed-State Self-Attention Network
	(QMSAN)~\cite{chen2025quantum} further enhanced quantum attention
	expressiveness through phase-preserving representations and mixed-state
	similarity computation, respectively.
	
	These purely quantum schemes, however, share a common constraint: the
	limited qubit counts of NISQ devices preclude direct encoding of
	high-dimensional image inputs. The hybrid paradigm of VQAs offers a
	natural remedy---employing PQCs for quantum computation while
	delegating high-dimensional input preprocessing to classical networks,
	with both components cooperating to perform attention.
	Mari et al.~\cite{mari2020transfer} pioneered this approach, using
	pretrained classical networks as feature extractors to achieve
	high-resolution image classification on real quantum hardware.
	HQViT~\cite{zhang2025hqvit} and Quantum Discrete Fourier Transform-Based 
	Hierarchical Self-Attention (QHSA-ViT)~\cite{quQHSAViTQuantumDiscrete2025} further advanced the
	hybrid paradigm to vision Transformers, computing attention via quantum
	swap tests and quantum discrete Fourier transforms respectively,
	alleviating the high-resolution modeling bottleneck while remaining
	NISQ-compatible. The Quantum Adaptive Self-Attention
	(QASA)~\cite{chen2025quantumb} demonstrated that introducing only 36
	quantum parameters into a single encoder layer suffices to outperform
	fully quantum distributional models on certain time-series prediction
	tasks.
	
	Most closely related to our design are Tomal
	et al.~\cite{tomal2025quantum} and Smaldone
	et al.~\cite{smaldone2025hybrid}, both replacing the query--key dot
	product with quantum circuits to compute attention scores. While Tomal
	et al. apply quantum circuits as entanglement-aware kernel similarity
	for NLP tasks, Smaldone et al. aim to reduce the time complexity of
	feature dimensions by replicating the dot-product operation via quantum
	states, targeting molecular generation. Neither systematically
	investigates PQCs as a general-purpose scoring function on vision
	tasks---the precise gap this work addresses.
	
	The development of classical attention mechanisms reveals the
	centrality of the scoring function in attention design. From additive
	to dot-product formulations, the choice of scoring function
	significantly shapes model performance;
	yet the dot product's dominance stems from computational efficiency rather than theoretical uniqueness.
	VQAs provide a computational framework combining theoretical advantages
	with near-term hardware feasibility, offering a principled path toward
	exploring beyond classical scoring functions.
	
	\begin{figure}[!t]
		\centering
		\includegraphics[width=\columnwidth]{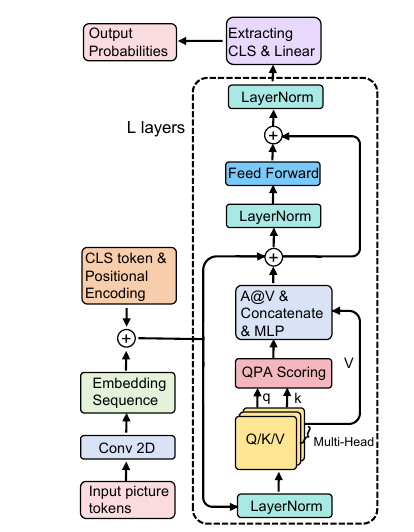}
		\caption{Overall architecture of \mbox{QPSAN}. The core distinction
			from classical \mbox{ViT} is the replacement of the dot-product operation with the \mbox{QPA Scoring Module}.}
		\label{fig:over_structure}
	\end{figure}
	
	\section{The Proposed Method}
	\label{sec:proposed_method}
	
	\subsection{Overview}
	\label{sec:overview}
	
	As illustrated in Fig.~\ref{fig:over_structure}, QPSAN builds on the
	ViT~\cite{dosovitskiy2020image} architecture, replacing the dot-product
	scoring function in multi-head self-attention with QPA. Implemented via
	a two-qubit PQC, QPA establishes nonlinear interactions between queries
	and keys through a quantum entanglement structure, with only 5
	trainable parameters
	($\theta_{\text{s}}, \gamma_{\text{d}}, \gamma_{\text{s}}, \alpha, \beta$).
	
	Input images are partitioned into non-overlapping patches based on
	patch size and mapped to input vectors via convolutional patch
	embedding, then concatenated with a learnable CLS token and
	positional encodings to form a token sequence.
	This sequence passes through $L$ Transformer encoder layers---each
	comprising Pre-Norm, multi-head self-attention, residual connections,
	and a feed-forward network---from which the CLS token representation is
	extracted and projected to classification logits via a linear head.
	
	Q, K, and V are generated through linear projection and
	multi-head splitting, following the same attention weight computation
	pipeline as ViT; the core distinction lies in how the attention score
	matrix $\bm{A}$ is generated. ViT employs scaled dot products,
	producing scores with an unbounded range
	$(-\infty,+\infty)$~\cite{henry2020query}. QPSAN, by contrast,
	encodes each dimension of Q and K into the two-qubit circuit,
	computing per-dimension attention scores via joint measurement. The
	output is naturally bounded in $[0,1]$, requiring no scaling.
	Per-dimension scores are then summed and passed through Softmax
	normalization to yield the final attention weights. Circuit
	architecture and the attention computation pipeline are detailed in
	Section~\ref{subsec:quantum_parametric_attention_circuit}; theoretical
	properties are analyzed in Section~\ref{subsec:properties}.
	
	\subsection{QPA DESIGN}
	\label{subsec:quantum_parametric_attention_circuit}
	
\begin{figure}[!t]
		\centering
		\includegraphics[width=\columnwidth]{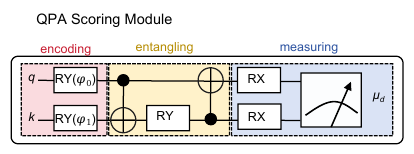}
		\caption{\mbox{QPA} module circuit structure. The upper wire is
			\mbox{Q0}, corresponding to the $d$-th dimension input scalar
			$q = Q_{i,d}$; the lower wire is \mbox{Q1}, corresponding to the $d$-th dimension
			input scalar $k = K_{j,d}$; the circuit runs independently for each dimension,
			outputting per-dimension attention score components $\mu_d$ via
			joint measurement.}
		\label{fig:quantum_circuit}
	\end{figure}

	Prior work demonstrates that two-qubit entangling circuits suffice to
	capture feature correlations inaccessible to classical
	networks~\cite{zhang2025dual,zhaoEntanglementinducedProvableRobust2025}.
	Motivated by this, we adopt the alternating-layer design of the Quantum
	Approximate Optimization Algorithm (QAOA) to construct a two-qubit PQC
	for attention scoring. Standard QAOA builds variational circuits by
	alternating Problem Hamiltonian layers and Mixer Hamiltonian layers;
	analogously, as shown in Fig.~\ref{fig:quantum_circuit}, we employ a
	serial encode--entangle--measure structure, with circuit outputs serving
	directly as per-dimension attention score components, summed across $D$
	dimensions (computed dimension-by-dimension over all $d_h$ dimensions,
	but aggregating only the first $D$) to obtain the complete attention
	score.
	
	\subsubsection{Encoding}
	
	The encoding stage comprises three sequential RY-gate rotations,
	applied to each query--key pair in turn.
	
	\textit{Step~1---Initial Encoding.} A $\pi/4$ offset is introduced
	to avoid the linear response region of the RY gate:
	\begin{equation}
		U_{\text{init}}(q, k) = \mathrm{RY}\!\left(\frac{\pi}{4} +
		\theta_{\text{s}} q\right) \otimes \mathrm{RY}\!\left(\frac{\pi}{4}
		+ \theta_{\text{s}} k\right).
	\end{equation}
	
	This $\pi/4$ offset avoids the linear response region of the RY gate;
	under two-qubit interference it accumulates to $\pi/2$ (see Lemma~2),
	mapping inputs onto orthogonal sine and cosine bases and thereby
	breaking the pure even symmetry of the output.
	
	\textit{Step~2---Difference Encoding.} Opposing rotations capture
	the degree of difference between the query and key:
	\begin{equation}
		U_{\text{diff}}(q, k) = \mathrm{RY}\!\left(\gamma_{\text{d}}
		(q - k)\right) \otimes \mathrm{RY}\!\left(-\gamma_{\text{d}}
		(q - k)\right).
	\end{equation}
	
	\textit{Step~3---Sum Encoding.} Co-directional rotations capture
	the overall activation magnitude:
	\begin{equation}
		U_{\text{sum}}(q, k) = \mathrm{RY}\!\left(\gamma_{\text{s}}
		(q + k)\right) \otimes \mathrm{RY}\!\left(\gamma_{\text{s}}
		(q + k)\right).
	\end{equation}
	
	By the additivity of RY gates,
	$\mathrm{RY}(\alpha)\mathrm{RY}(\beta) = \mathrm{RY}(\alpha+\beta)$,
	the three steps collapse into a single equivalent RY layer:
	\begin{equation}
		U_{\text{enc}}(q, k) = \mathrm{RY}(\varphi_{0}) \otimes
		\mathrm{RY}(\varphi_{1}),
	\end{equation}
	where
	\begin{align}
		\varphi_{0} &= \frac{\pi}{4} + (\theta_{\text{s}} + \gamma_{\text{d}}
		+ \gamma_{\text{s}})q + (\gamma_{\text{s}} - \gamma_{\text{d}})k, \\
		\varphi_{1} &= \frac{\pi}{4} + (\gamma_{\text{s}} - \gamma_{\text{d}})q
		+ (\theta_{\text{s}} + \gamma_{\text{d}} + \gamma_{\text{s}})k.
	\end{align}
	In the collapsed form, the $q$ and $k$ coefficients in $\varphi_0$ and
	$\varphi_1$ are exactly swapped. The three-step formulation makes the
	design rationale explicit; the collapsed form is more compact. Both are
	mathematically equivalent.
	
	Before entanglement, the encoding stage already incorporates the
	correlation between $q$ and $k$ into the rotation angles via linear
	combinations $(q-k)$ and $(q+k)$. Using independent encoding
	$\mathrm{RY}(\theta q) \otimes \mathrm{RY}(\theta k)$ instead would
	leave the two-qubit state fully separable prior to entanglement,
	confining query--key interaction entirely to the entanglement layer.
	Empirically, training with independent encoding leads to a significant
	performance drop.
	
	\subsubsection{Entangling}
	
	Following encoding, the entanglement stage applies bidirectional CNOTs
	flanking a parameterized rotation:
	\begin{equation}
		\scalebox{0.92}{$U_{\text{ent}}(q, k) = \mathrm{CNOT}_{1\to0}
			\cdot \left(I \otimes \mathrm{RY}\left(\alpha(q+k)\right)\right)
			\cdot \mathrm{CNOT}_{0\to1}$},
	\end{equation}
	where $\mathrm{RY}(\alpha(q+k))$ acts solely on the second qubit,
	allowing the entanglement strength to adapt to the input $(q+k)$
	rather than applying a fixed entangling operation.
	
	The Mixer layer then applies identical RX rotations to both qubits to
	mix quantum states:
	\begin{equation}
		U_{\text{mix}}(\beta) = \mathrm{RX}(2\beta) \otimes
		\mathrm{RX}(2\beta).
	\end{equation}
	
	The complete circuit evolution is:
	\begin{equation}
		|\psi(q, k)\rangle = U_{\text{mix}}(\beta) \cdot U_{\text{ent}}
		(q, k) \cdot U_{\text{enc}}(q, k)\, |00\rangle.
	\end{equation}
	
	\subsubsection{Measuring}
	
	\begin{definition}[Per-Dimension Quantum Attention Score]
		Let the $d$-th dimension query and key components form a two-qubit
		product state after angle encoding. The per-dimension attention score
		is defined via joint measurement in the computational basis as
		\begin{equation}
			\mu_d(q,k) = P(|00\rangle) + P(|11\rangle),
		\end{equation}
		where $\mu_d(q,k) \in [0,1]$ is the sum of probabilities of both
		qubits occupying the same basis state, reflecting the alignment
		between the query and key components in the computational basis.
		For a two-qubit system, there are six possible joint-measurement
		outcomes: four depend on a single qubit's state alone and cannot
		capture the joint interaction of $(q,k)$. The remaining
		two---$P(|00\rangle)+P(|11\rangle)$ and
		$P(|01\rangle)+P(|10\rangle)$---sum to exactly~1 and mutually
		determine each other. We adopt the former.
	\end{definition}
	
	The circuit contains 5 trainable parameters in total: initial encoding
	scale $\theta_{\text{s}}$ (initialized to $0.5$), difference encoding
	strength $\gamma_{\text{d}}$, sum encoding strength $\gamma_{\text{s}}$,
	entanglement strength $\alpha$, and Mixer angle $\beta$; the parameters
	$\gamma_{\text{d}}, \gamma_{\text{s}}, \alpha, \beta$ are randomly
	initialized from $\mathcal{N}(0, 0.1^2)$. All parameters are shared
	across all query--key pairs within the same Transformer layer and
	updated via backpropagation; different layers maintain independent
	parameter sets.
	
	The scoring function $\mu_{d}(q,k)$ output by the circuit forms the
	core of QPA. Within the multi-head self-attention framework, QPA
	replaces the classical dot product: for each token pair $(Q_i, K_j)$,
	the scalar components $(Q_{i,d}, K_{j,d})$ are fed into the quantum
	circuit dimension by dimension; encoding, evolution, and joint
	measurement yield per-dimension scores $\mu_d \in [0,1]$, which are
	then summed across $D=16$ dimensions to form the attention score
	\begin{equation}
		[\bm{A}]_{ij} = \sum_{d=1}^{D} \mu_d(Q_{i,d}, K_{j,d}).
	\end{equation}
	After Softmax normalization, the result is combined with the Value
	matrix via weighted summation:
	\begin{equation}
		\text{Attention}(\bm{Q}, \bm{K}, \bm{V}) =
		\operatorname{softmax}(\bm{A})\bm{V},
	\end{equation}
	where $\operatorname{softmax}(\cdot)$ normalizes along the column
	dimension (i.e., over each row $[\bm{A}]_{i,:}$). Unlike the classical
	dot product $\bm{Q}\bm{K}^\top$, which naturally aggregates all
	dimensions at the vector level, the quantum circuit operates on scalar
	pairs and requires independent per-dimension computation followed by
	aggregation. Experiments confirm that $D=16$ represents the optimal
	trade-off between efficiency and performance.
	
	\subsection{QPA Properties}
	\label{subsec:properties}
	
	This section derives the analytical structure of the attention score
	$\mu$ and establishes its fundamental properties: Lemmas~1--2 analyze
	the structural characteristics of the three-step encoding,
	Properties~1--3 establish the fundamental properties of $\mu$, and
	Theorems~1--2 quantify the effective degrees of freedom of the
	encoding layer. Complete proofs of all results are provided in the
	Supplementary Material.
	
	\begin{lemma}[Non-separable kernel]
		\label{lem:nonsep}
		Define
		\begin{equation}
			\lambda_1 \triangleq \theta_{\text{s}} + \gamma_{\text{d}} + \gamma_{\text{s}} ,
			\quad
			\lambda_2 \triangleq \gamma_{\text{s}} - \gamma_{\text{d}} .
		\end{equation}
		Here $\lambda_1$ is the self-coefficient (Qubit~0 senses $q$,
		Qubit~1 senses $k$) and $\lambda_2$ is the cross-coefficient
		(Qubit~0 senses $k$, Qubit~1 senses $q$), allowing each qubit to
		sense the input feature of the other.
		
		The following analysis is based on the product state output of the
		three-step encoding layer, prior to the entanglement layer; once the
		entanglement layer acts, the quantum state is no longer a product
		state, and the separability analysis of $K_{\text{enc-3}}$ does not
		extend to the full circuit. The quantum kernel
		$K(\bm{x}_1,\bm{x}_2) = |\langle\psi(\bm{x}_1)|\psi(\bm{x}_2)\rangle|^2$
		then simplifies to
		\begin{equation}
			K_{\text{enc-3}} =
			\cos^2\!\left(\lambda_1'\Delta q+\lambda_2'\Delta k\right)
			\cdot
			\cos^2\!\left(\lambda_2'\Delta q+\lambda_1'\Delta k\right) ,
			\label{eq:Kenc3}
		\end{equation}
		where $\lambda_1' \triangleq \lambda_1/2$,
		$\lambda_2' \triangleq \lambda_2/2$,
		$\Delta q = q_2-q_1$, and \mbox{$\Delta k = k_2-k_1$.}
		When $\lambda_1\neq 0$ and $\lambda_2\neq 0$, $K_{\text{enc-3}}$ is
		non-separable, i.e., it cannot be decomposed into the form
		\mbox{$f(\Delta q)\cdot g(\Delta k)$.} By contrast, single-parameter
		independent encoding
		$\mathrm{RY}(\pi/4+Eq)\otimes\mathrm{RY}(\pi/4+Ek)$
		($E\triangleq\theta_{\text{s}}$) degenerates to the separable kernel
		\begin{equation}
			K_{\text{enc-1}} = \cos^2\!\left(\frac{E\Delta q}{2}\right)\cdot
			\cos^2\!\left(\frac{E\Delta k}{2}\right) .
		\end{equation}
		A separable kernel's responses to $\Delta q$ and $\Delta k$ are
		mutually independent---changes in $\Delta q$ do not affect
		$g(\Delta k)$, and changes in $\Delta k$ do not affect
		$f(\Delta q)$---and thus cannot capture interaction patterns arising
		from the joint variation of $q$ and $k$. The non-separability of
		$K_{\text{enc-3}}$ enables the three-step encoding's feature space to
		sense the joint variation of $(q,k)$ rather than sensing each axis
		independently along $\Delta q$ and $\Delta k$.
	\end{lemma}
	
	\begin{lemma}[Independent frequency modulation]
		\label{lem:two-freq}
		Considering the encoding layer only (product state), the QPA scoring
		function is
		\begin{equation}
			\mu(q,k) = \frac{1}{2}
			+\frac{1}{4}\cos\!\big(\omega_{\text{d}}(q-k)\big)
			-\frac{1}{4}\sin\!\big(\omega_{\text{s}}(q+k)\big) ,
			\label{eq:mu-two-freq}
		\end{equation}
		where $\omega_{\text{d}} = \theta_{\text{s}}+2\gamma_{\text{d}}$ and
		$\omega_{\text{s}} = \theta_{\text{s}}+2\gamma_{\text{s}}$.
		The two frequency components lie along the $(q-k)$ and $(q+k)$
		directions, respectively: $\gamma_{\text{d}}$ modulates only
		$\omega_{\text{d}}$, $\gamma_{\text{s}}$ modulates only
		$\omega_{\text{s}}$, and the two can be optimized independently;
		by contrast, single-parameter encoding locks both directional
		frequencies to the same parameter $\theta_{\text{s}}$, making
		separate adjustment impossible.
	\end{lemma}
	
	\begin{property}[Boundedness]
		\label{prop:bounded}
		For any input pair $(q,k)\in\mathbb{R}^2$ and any circuit parameters,
		the QPA scoring function satisfies $\mu(q,k)\in[0,1]$.
	\end{property}
	
	\begin{property}[Non-symmetry of the full circuit, $\alpha\neq 0$]
		\label{prop:nonsym}
		Except on a degenerate parameter set of measure zero in the parameter
		space, for the full QPA circuit under general parameters
		($\alpha\neq 0$, $\lambda_1\neq\lambda_2$), the scoring function
		satisfies $\mu(q,k)\neq\mu(k,q)$.
	\end{property}
	
	\begin{property}[Non-monotonicity of the full circuit
		over $\mathbb{R}^2$]
		\label{prop:nonmono}
		Under general parameters (excluding a degenerate set of measure zero
		in the parameter space), the scoring function $\mu(q,k)$ of the full
		QPA circuit over input domain $\mathbb{R}^2$ is not monotone with
		respect to $|q-k|$.
	\end{property}
	
	\begin{theorem}[Effective degrees of freedom of the encoding layer]
		\label{thm:dof-enc}
		The 3 trainable parameters $(\theta_{\text{s}}, \gamma_{\text{d}},
		\gamma_{\text{s}})$ of the QPA encoding layer contribute only 2
		independent degrees of freedom to the scoring function.
	\end{theorem}
	
	\begin{theorem}[Bounds on the effective degrees of freedom of the
		full circuit]
		\label{thm:dof_full}
		The effective dimension of the scoring function class of the full QPA
		circuit is defined as
		\begin{equation}
			d_Q \triangleq \max_{\bm{\Theta}\in\mathbb{R}^5}
			\operatorname{rank}(\bm{J}_{\text{full}}(\bm{\Theta})) ,
		\end{equation}
		where $\bm{\Theta}=(\theta_{\text{s}},\gamma_{\text{d}},
		\gamma_{\text{s}},\alpha,\beta)$ and $\bm{J}_{\text{full}}$ is the
		Jacobian matrix of the full circuit output with respect to
		$\bm{\Theta}$. Then
		\begin{equation}
			2 \leq d_Q \leq 4 .
		\end{equation}
	\end{theorem}
	
	\begin{remark}[Structural constraints and parameter efficiency]
		\label{rem:struct}
		The physical origin of Theorem~2 lies in the RY-gate additivity of
		the three-step encoding: the rotation angles applied by each step on
		the same qubit superpose linearly, so the 3 independent parameters
		$(\theta_{\text{s}},\gamma_{\text{d}},\gamma_{\text{s}})$ ultimately
		fold into 2 effective frequencies $(\omega_{\text{d}},\omega_{\text{s}})$;
		together with the entanglement layer parameter $\alpha$ and the Mixer
		layer parameter $\beta$, each contributing 1 independent degree of
		freedom, the effective dimension of the full circuit is
		upper-bounded by~4. This structural constraint implies that, regardless
		of how $(\theta_{\text{s}},\gamma_{\text{d}},\gamma_{\text{s}},
		\alpha,\beta)$ are adjusted, the dimension of the scoring function
		class expressible by QPA does not exceed~4; classical MLPs, by
		contrast, admit no such folding mechanism---each parameter
		independently modulates the output through nonlinear activations, and
		adding parameters typically expands the function class boundary---yet
		as the ablation experiments in
		Section~\ref{subsubsec:ablation_study} demonstrate, expanding
		expressiveness does not automatically translate into improved
		generalization.
	\end{remark}
	
	Subject to valid probability outputs (the boundedness of Property~1),
	the QPA scoring function $\mu(q,k)$ possesses three structural
	properties that distinguish it from classical dot-product attention.
	First, Lemmas~1--2 show that the three-step encoding layer (product
	state) produces a non-separable kernel when $\lambda_1\neq 0$ and
	$\lambda_2\neq 0$, with the frequencies $\omega_{\text{d}}$ and
	$\omega_{\text{s}}$ along the $(q-k)$ and $(q+k)$ directions
	independently adjustable by $\gamma_{\text{d}}$ and
	$\gamma_{\text{s}}$; together these endow the encoding layer with the
	potential to sense the joint variation of $(q,k)$, a capability absent
	from single-parameter encoding schemes. Second, Property~2 establishes
	the asymmetry of $\mu$: \mbox{$\mu(q,k)\neq\mu(k,q)$} means the scoring
	function can distinguish the directional roles of query and key---the
	same input pair directed from query to key versus from key to query may
	receive different attention weights, thereby capturing directional
	dependencies between tokens. This is a structural capability absent
	from classical dot-product attention, which is naturally symmetric
	under exchange of $q$ and $k$. Finally, Property~3 shows that $\mu$
	does not decay monotonically with $|q-k|$: in classical dot-product
	attention the scoring function typically decreases monotonically as the
	distance between query and key grows, naturally biasing the mechanism
	toward closer tokens; the QPA scoring function can in principle attain
	local maxima at arbitrary distances, giving the model the ability to assign 
	high weights to distant input
	pairs and thereby the potential to model non-local feature interactions. 
	Theorems~1 and~2 quantify the
	structural constraints: the 3 encoding parameters contribute only 2
	independent degrees of freedom, and the effective dimension of the
	full circuit satisfies \mbox{$2\leq d_Q\leq 4$}, providing one theoretical
	explanation for the failure of increased classical MLP parameters to
	improve performance observed in the ablation experiments.
	Properties~2 and~3 theoretically characterize the additional
	structural properties of the QPA scoring function that distinguish it
	from classical dot-product attention; their specific mechanisms of
	action await verification in future work.
	
	\section{Experiments}
	\label{sec:experiments}
	
	\subsection{Setup}
	\label{subsec:experimental_setup}
	
	All experiments are conducted on a single NVIDIA RTX 4090 GPU, with
	models implemented using the PyTorchQuantum
	library~\cite{wang2022torchquantum} and trained from scratch without
	pretrained weights. Training uniformly employs stochastic gradient
	descent (SGD) with momentum $0.9$ and zero weight decay; the learning
	rate decays from its initial value to near zero via cosine annealing
	over 100 epochs, with a 3-epoch warmup and a patience of 20 epochs.
	
	The four datasets span visual binary classification tasks of varying
	difficulty. FashionMNIST~(FM)~\cite{xiao2017fashion} offers relatively
	clear inter-class visual differences, serving as a low-complexity
	baseline. DirtyMNIST~(DM)~\cite{mukhoti2023deep} explicitly introduces
	label noise and epistemic uncertainty, assessing model performance under
	noisy and uncertain conditions. \mbox{CIFAR-10~(CF)}~\cite{krizhevsky2009learning}
	provides multi-channel RGB images, probing the model's capacity to
	handle multi-channel RGB inputs. FER2013~(FER)~\cite{goodfellow2013challenges}
	is a grayscale facial expression dataset with inherently ambiguous
	inter-class boundaries, evaluating model performance where class
	separability is intrinsically limited. Binary classification task
	selection and dataset configurations are detailed in
	Table~\ref{tab:dataset_setup}.
	
	\begin{table}[!t]
		\centering
		\caption{Dataset Configuration for Hyperparameter Search and Model Training}
		\label{tab:dataset_setup}
		\renewcommand{\arraystretch}{1.2}
		\begin{small}
			\begin{tabular}{>{\centering\arraybackslash}p{0.8cm}
					>{\centering\arraybackslash}p{2.2cm}
					>{\centering\arraybackslash}p{1.8cm}
					>{\centering\arraybackslash}p{1.8cm}}
				\toprule[1pt]
				Dataset & Classification Task & Hyperpar. (train/valid) & Training (train/valid) \\
				\midrule[0.5pt]
				FM  & T-shirt/top vs Trouser & 250/100 & 250/100  \\
				DM  & Number 4 vs Number 9   & 250/100 & 1000/300 \\
				CF  & Airplane vs Automobile & 250/100 & 500/200  \\
				FER & Happy vs Surprise      & 250/100 & 500/200  \\
				\bottomrule[1pt]
			\end{tabular}
		\end{small}
		\begin{tablenotes}
			\footnotesize
			\item Binary classification task categories for the four datasets,
			along with training and validation set sizes used in the
			hyperparameter search phase and the formal training phase.
		\end{tablenotes}
	\end{table}
	
	\begin{table}[!t]
		\centering
		\caption{Architectural Hyperparameters Across Datasets}
		\label{tab:hyperparameter_config}
		\renewcommand{\arraystretch}{1.2}
		\begin{small}
			\begin{tabular}{>{\centering\arraybackslash}p{0.8cm}
					>{\centering\arraybackslash}p{1.1cm}
					>{\centering\arraybackslash}p{1.1cm}
					>{\centering\arraybackslash}p{1.1cm}
					>{\centering\arraybackslash}p{1.1cm}
					>{\centering\arraybackslash}p{1.1cm}}
				\toprule[1pt]
				Dataset & Patch Size & Num. Layers & Num. Heads & Hidden Size & MLP Params \\
				\midrule[0.5pt]
				FM  & $7\times7$ & 1 & 3 & 192 & 0.3M \\
				DM  & $7\times7$ & 1 & 3 & 192 & 0.3M \\
				CF  & $8\times8$ & 4 & 4 & 256 & 2M   \\
				FER & $8\times8$ & 2 & 6 & 192 & 0.7M \\
				\bottomrule[1pt]
			\end{tabular}
		\end{small}
		\begin{tablenotes}
			\footnotesize
			\item Architectural hyperparameters for the four datasets,
			including patch size, number of attention layers, number of
			attention heads, hidden dimension, and MLP parameter count.
		\end{tablenotes}
	\end{table}
	
	All models are optimized with cross-entropy loss:
	\begin{equation}
		\mathcal{L} = -\frac{1}{N}\sum_{i=1}^{N}\log
		\frac{\exp(z_{i,y_i})}{\sum_{j=1}^{C}\exp(z_{i,j})},
	\end{equation}
	where $\bm{z}_i \in \mathbb{R}^C$ is the logit vector of sample $i$,
	$z_{i,j}$ its $j$-th component, $y_i$ the ground-truth label, $N$ the
	batch size, and $C$ the number of classes.
	
	\subsection{Metrics}
	\label{subsec:evaluation_metrics}
	
	For binary classification, this paper selects the metrics shown in
	Table~\ref{tab:evaluation_metrics} for evaluation.
	
	Since quantum and classical models share the same random seed pool,
	each independent run corresponds to identical data splits and
	initialization conditions, enabling pairwise matching of results. To
	eliminate the effect of randomness, all experiments are assessed via
	paired $t$-tests at significance level $\alpha = 0.05$; both one-tailed
	and two-tailed results are reported to ensure inferential consistency
	under directional and non-directional hypotheses alike, with Cohen's
	$d$ as the effect size measure. The null hypothesis is
	$H_0\colon \bar{p}_Q = \bar{p}_C$, with one-tailed alternative
	$H_1\colon \bar{p}_Q > \bar{p}_C$ and two-tailed alternative
	$H_1\colon \bar{p}_Q \neq \bar{p}_C$, where $\bar{p}_Q$ and
	$\bar{p}_C$ denote the mean validation accuracy of the quantum and
	classical models, respectively.
	
	\begin{table}[!t]
		\centering
		\caption{Evaluation Metrics}
		\label{tab:evaluation_metrics}
		\renewcommand{\arraystretch}{1.5}
		\begin{small}
			\begin{tabular}{>{\centering\arraybackslash}p{1.5cm}
					>{\centering\arraybackslash}p{3.2cm}
					>{\raggedright\arraybackslash}p{3.2cm}}
				\toprule[1pt]
				Metric & Formula & Description \\
				\midrule[0.5pt]
				Accuracy &
				$\dfrac{\mathit{TP} + \mathit{TN}}{\mathit{TP} + \mathit{TN} + \mathit{FP} + \mathit{FN}}$ &
				Reflects overall classification accuracy \\[2ex]
				Precision &
				$\dfrac{\mathit{TP}}{\mathit{TP}+\mathit{FP}}$ &
				Proportion of truly positive samples among all predicted
				positives \\[2ex]
				Recall &
				$\dfrac{\mathit{TP}}{\mathit{TP}+\mathit{FN}}$ &
				Proportion of actual positives correctly identified \\[2ex]
				F1 Score &
				$2 \times \dfrac{\text{Precision} \times \text{Recall}}{\text{Precision} + \text{Recall}}$ &
				Harmonic mean of Precision and Recall \\[2ex]
				AUC-ROC &
				Area under the ROC curve &
				Measures the model's capacity to distinguish positive from
				negative samples \\
				\bottomrule[1pt]
			\end{tabular}
		\end{small}
		\begin{tablenotes}
			\footnotesize
			\item $\mathit{TP}$, $\mathit{TN}$, $\mathit{FP}$, and
			$\mathit{FN}$ denote the counts of true positives, true
			negatives, false positives, and false negatives, respectively.
			AUC-ROC denotes the area under the ROC curve, abbreviated as
			AUC where no ambiguity arises.
		\end{tablenotes}
	\end{table}
	
	\subsection{Hyperparameter Tuning}
	\label{subsubsec:hyperparameter_tuning}
	
	We conduct a grid search over learning rate~(lr) and batch size~(bs)
	across 20 hyperparameter combinations, each trained independently on 5
	different random seeds with the mean taken as the evaluation metric.
	Regarding the composition of the DM dataset: its training and
	validation splits each contain 50\% AmbiguousMNIST samples, which
	introduce structural image blur and label uncertainty to simulate a
	noisy real-world learning environment.
	
	\begin{figure*}[!t]
		\centering
		\includegraphics[width=\linewidth]{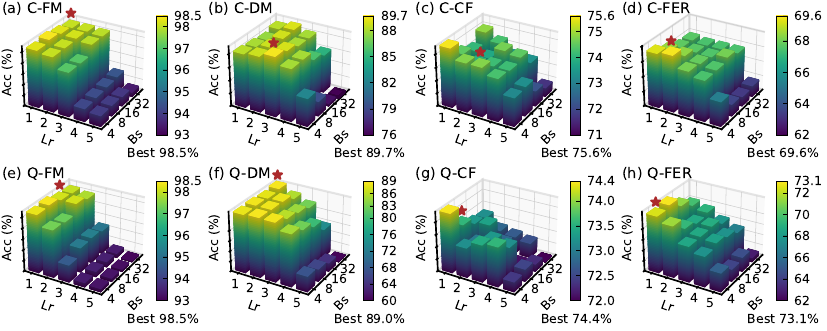}
		\caption{Bar chart of hyperparameter search results. 
			\mbox{C} denotes the classical \mbox{ViT} model; \mbox{Q} 
			denotes the quantum \mbox{QPSAN} model. 
			Indices 1--5 on the learning rate axis correspond to learning rates
			$5\times10^{-3}$, $2\times10^{-3}$, $10^{-3}$,
			$5\times10^{-4}$, and $10^{-4}$, respectively; the final training
			configuration is determined jointly by accuracy and training
			stability (measured by standard deviation), marked with a red star.
			Panels~(c) and~(g) show CF dataset search results: the combination
			with the highest mean accuracy has a relatively large standard
			deviation (\mbox{C-CF}: $0.012$; \mbox{Q-CF}: $0.011$), so the
			configuration with slightly lower accuracy but substantially smaller
			standard deviation (\mbox{C-CF}: $0.005$; \mbox{Q-CF}: $0.0004$)
			is selected instead.}
		\label{fig:lr_bs_search}
	\end{figure*}
	
	The search results (Fig.~\ref{fig:lr_bs_search}) show that the
	learning rate dominates model performance, with batch size exerting far
	less influence. The peak mean accuracy of the quantum model across
	multiple datasets consistently occurs at $5\times10^{-3}$, suggesting
	that quantum models may require larger learning rates to adequately
	explore the parameter space. On FM, DM, and FER, the quantum model is
	more sensitive to the learning rate than its classical counterpart,
	with the DM quantum model spanning an accuracy range of 25 percentage
	points \mbox{($64\%$--$89\%$)}; CF exhibits the smallest accuracy fluctuation
	among the four datasets, with mean accuracy variations of only $3.9\%$
	and $2.4\%$ for the classical and quantum models, respectively.
	
	\section{Results and Discussion}
	\label{sec:results_and_analysis}
	\begin{figure*}[!t]
		\centering
		\includegraphics[width=\linewidth]{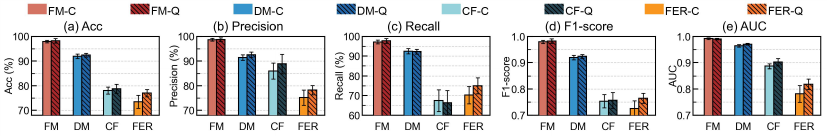}
		\caption{Comparison of \mbox{QPSAN} and \mbox{ViT} results on the four
			datasets across five evaluation metrics: validation accuracy,
			precision, recall, F1-score, and \mbox{AUC-ROC}.}
		\label{fig:accuracy}
	\end{figure*}
	
	\subsection{Results}
	\label{subsec:results}
	
	Fig.~\ref{fig:accuracy} presents the performance of QPSAN and ViT
	across five evaluation metrics (complete numerical results are
	provided in the supplementary material). The most pronounced improvement appears
	on FER: the quantum model's mean accuracy exceeds the classical
	baseline by 3.54\% (from 73.45\% to 76.99\%), with standard deviation
	reduced from 2.67\% to 1.50\%---a 43.8\% decrease. The paired
	$t$-test yields one-tailed $p = 8.6\times10^{-6}$ and two-tailed
	$p = 1.7\times10^{-5}$, both far below $\alpha = 0.05$, with Cohen's
	$d = 1.27$ indicating a very large effect. DM and CF likewise show
	statistically significant improvements, with Cohen's $d$ of 0.55 and
	0.53, respectively---both medium effects (one-tailed $p < 0.05$). On
	FM, the accuracy difference is smaller and does not reach statistical
	significance ($p > 0.05$), with Cohen's $d = 0.37$ indicating a small
	effect. QPSAN outperforms ViT in mean accuracy across all four
	datasets, with the advantage more pronounced on datasets with blurrier
	class boundaries and more noise. Detailed accuracy and $t$-test 
	results are reported in Table~\ref{tab:ttest_results}; 
	analysis of the underlying causes of the performance differences 
	is provided in Section~\ref{subsec:visualization_and_interpretability_analysis}.
	
	\begin{table*}[!t]
		\centering
		\caption{Paired $t$-Test Results Across Datasets}
		\label{tab:ttest_results}
		\renewcommand{\arraystretch}{1.2}
		\begin{small}
			\begin{tabular}{>{\raggedright\arraybackslash}p{3.0cm}
					>{\centering\arraybackslash}p{3.2cm}
					>{\centering\arraybackslash}p{3.2cm}
					>{\centering\arraybackslash}p{3.2cm}
					>{\centering\arraybackslash}p{3.2cm}}
				\toprule[1pt]
				& FM & DM & CF & FER \\
				\midrule[0.5pt]
				Quantum Model Acc
				& $98.22\%\pm0.87\%$ & $92.39\%\pm0.74\%$
				& $78.81\%\pm1.83\%$ & $76.99\%\pm1.50\%$ \\
				Classical Model Acc
				& $97.93\%\pm0.54\%$ & $91.87\%\pm0.91\%$
				& $78.06\%\pm1.44\%$ & $73.45\%\pm2.60\%$ \\
				Acc Diff (Q$-$C)
				& $+0.30\%$ & $+0.53\%$ & $+0.75\%$ & $+3.54\%$ \\
				$p$ (one-tail)
				& $0.0552$ & $0.0113$ & $0.0138$ & $<0.0001$ \\
				$p$ (two-tail)
				& $0.1105$ & $0.0226$ & $0.0275$ & $<0.0001$ \\
				Cohen's $d$
				& $0.3744$ & $0.5548$ & $0.5338$ & $1.2744$ \\
				95\% CI
				& $[-0.0007,\ 0.0068]$ & $[0.0008,\ 0.0097]$
				& $[0.0009,\ 0.0141]$ & $[0.0224,\ 0.0484]$ \\
				$t$-statistic
				& $1.6742$ & $2.4810$ & $2.3875$ & $5.6990$ \\
				Significance
				& n.s. & * & * & *** \\
				\bottomrule[1pt]
			\end{tabular}
		\end{small}
		\begin{tablenotes}
			\footnotesize
			\item Acc Diff denotes the accuracy difference between the
			quantum and classical models (Q$-$C); 95\% CI represents the
			confidence interval of the mean difference; Cohen's $d$
			indicates the effect size.
			\item n.s.: $p > 0.05$;\quad *: $0.01 < p \leq 0.05$;\quad
			**: $0.001 < p \leq 0.01$;\quad ***: $p \leq 0.001$.
		\end{tablenotes}
	\end{table*}
	
	A notable contrast exists between the hyperparameter search phase
	(5~runs per combination) and the formal experiments: during the search,
	the quantum model's mean accuracy on DM and CF fell below the classical
	baseline by 0.70\% and 1.20\%, respectively; this deficit reversed
	upon expanding the dataset and increasing to 20~runs in the formal
	experiments. No such reversal occurred on FM or FER. This pattern suggests 
	that with limited sample size and run count, performance differences 
	of quantum models may be masked by stochastic variation---motivating 
	our adoption of 20~independent runs with paired $t$-tests to ensure 
	statistical reliability of the conclusions.
	
	\subsection{Ablation Study}
	\label{subsubsec:ablation_study}
	
	The ablation study proceeds from four hypotheses, each testing the
	contribution of a specific design factor to QPSAN's performance:
	\begin{enumerate}
		\renewcommand{\labelenumi}{(\arabic{enumi})}
		\item If nonlinear scoring functions contribute substantially,
		MLPSAN---which replaces the quantum circuit with an MLP---should
		match QPSAN.
		\item If bounded attention scores contribute substantially,
		CosViT---which naturally produces bounded attention scores---should
		match QPSAN.
		\item If the overall shift in attention paradigm contributes
		substantially, LinearViT---which equally changes the attention
		paradigm---should match QPSAN.
		\item If the structural design of the encoding layer
		(non-separable kernel and independent frequency modulation)
		contributes substantially, QPSAN-Ind---which degrades to
		single-parameter independent encoding---should show a significant
		performance gap relative to QPSAN.
	\end{enumerate}
	Should QPSAN retain a significant advantage over all the above
	controls, this suggests that the performance gains may stem from the
	overall structural inductive bias of the quantum circuit, rather than
	being explained by any single factor in isolation.
	
	For hypothesis~(1), the quantum circuit within the QPA module is
	replaced by small MLPs: a two-layer variant with 49 parameters
	(MLPSAN-49) and a three-layer variant with 585 parameters
	(MLPSAN-585). Let $\bm{W}^{(\ell)}$ and $\bm{b}^{(\ell)}$ denote the
	weight and bias of layer $\ell$ (with the final layer producing vector
	$\bm{w}^{(L)}$ and scalar $b^{(L)}$); intermediate layers use
	$\tanh(\cdot)$ as the activation function, and the final output is
	mapped to $[0,1]$ via $\operatorname{sigmoid}$. The feature
	construction vector is \mbox{$\bm{f}(q,k) = [q,\, k,\, q{-}k,\,
		q{+}k]^\top$}, consistent with the encoding input of the quantum
	circuit.
	
	Denoting $\bm{h}^{(1)} = \tanh\!\left(\bm{W}^{(1)} \bm{f}(q,k) +
	\bm{b}^{(1)}\right)$, the scoring function of MLPSAN-49 (two layers)
	is:
	\begin{equation}
		S_d(q, k) = \sigma\!\left({\bm{w}^{(2)}}^\top
		\bm{h}^{(1)} + b^{(2)}\right).
	\end{equation}
	Denoting $\bm{h}^{(2)} = \tanh\!\left(\bm{W}^{(2)} \bm{h}^{(1)} +
	\bm{b}^{(2)}\right)$, the scoring function of MLPSAN-585 (three
	layers) is:
	\begin{equation}
		S_d(q, k) = \sigma\!\left({\bm{w}^{(3)}}^\top
		\bm{h}^{(2)} + b^{(3)}\right).
	\end{equation}
	The parameter matrix dimensions are as follows: \mbox{MLPSAN-49} uses
	$\bm{W}^{(1)}\in\mathbb{R}^{8\times4}$ and
	$\bm{w}^{(2)}\in\mathbb{R}^{8}$; MLPSAN-585 uses
	$\bm{W}^{(1)}\in\mathbb{R}^{64\times4}$,
	$\bm{W}^{(2)}\in\mathbb{R}^{4\times64}$, and
	$\bm{w}^{(3)}\in\mathbb{R}^{4}$.
	
	For hypothesis~(2), CosViT is introduced as a baseline: built on
	Scaled Cosine Attention~\cite{liu2021swin}, it employs a cosine
	attention mechanism with naturally bounded scores, where $\tau$ is a
	learnable temperature parameter capped at 100. The scoring function is:
	\begin{equation}
		S_d(\bm{q}, \bm{k}) = \frac{\bm{q}}{|\bm{q}|_2} \cdot
		\frac{\bm{k}^\top}{|\bm{k}|_2} \cdot
		\exp\!\big(\min(\log\tau,\;\log 100)\big),
	\end{equation}
	where $\tfrac{\bm{q}}{|\bm{q}|_2}\cdot\tfrac{\bm{k}^\top}{|\bm{k}|_2}$
	is the cosine similarity between $\bm{q}$ and $\bm{k}$, with range
	$[-1,1]$; CosViT uses the same aggregation as standard ViT.
	
	For hypothesis~(3), LinearViT is introduced as a baseline: built on
	the Linear Transformer~\cite{katharopoulos2020transformers}, it
	replaces $\operatorname{softmax}$ with kernel mapping and matrix
	associativity, where \mbox{$\phi(\cdot)=\mathrm{elu}(\cdot)+1$} is the
	kernel mapping function and $\epsilon$ is a numerical stability
	constant. The attention mechanism is:
	\begin{equation}
		\text{Attention}(\bm{Q},\bm{K},\bm{V}) =
		\frac{\phi(\bm{Q})\bigl(\phi(\bm{K})^\top \bm{V}\bigr)}
		{\phi(\bm{Q})^\top\textstyle\sum_j\phi(\bm{k}_j)+\epsilon}.
	\end{equation}
	LinearViT exploits matrix associativity to compute
	$\phi(\bm{K})^\top \bm{V}\in\mathbb{R}^{d_h\times d_h}$ first, then
	left-multiplies by $\phi(\bm{Q})$, reducing attention computation from
	$O(N^2 d_h)$ to $O(Nd_h^2)$ without introducing additional learnable
	parameters.
	
	For hypothesis~(4), the ablation model QPSAN-Ind is constructed: the
	entangling layer, Mixer layer, and joint measurement structure of the
	full circuit are retained; only the three-step encoding is replaced by
	single-parameter independent encoding:
	\begin{equation}
		U_{\text{Ind}}(q,k) =
		\mathrm{RY}\!\left(\frac{\pi}{4} + \theta_{\text{s}} q\right)
		\otimes
		\mathrm{RY}\!\left(\frac{\pi}{4} + \theta_{\text{s}} k\right).
	\end{equation}
	In this encoding scheme, each qubit senses only its own input ($q$ or
	$k$); the corresponding encoding kernel degrades to the separable
	kernel $K_{\text{enc-1}}$ in Lemma~1, and the frequencies along both
	directions are determined by the same parameter $\theta_{\text{s}}$,
	precluding the independent frequency modulation described in Lemma~2.
	
	QPSAN, MLPSAN, and QPSAN-Ind share the same aggregation scheme:
	summing over the first $D$ dimensions followed by
	$\operatorname{softmax}$ normalization (see
	Section~\ref{subsec:quantum_parametric_attention_circuit}).
	
	Following hyperparameter search, the optimal configurations are:
	MLPSAN-49 ($lr=5\times10^{-3},\ \mbox{bs=4}$), \mbox{MLPSAN-585}
	($lr=2\times10^{-3},\ \mbox{bs=4}$), CosViT \mbox{($lr=1\times10^{-3},\ bs=32$)},
	LinearViT ($lr=5\times10^{-3},\ bs=4$), and QPSAN-Ind
	($lr=5\times10^{-3},\ \mbox{bs=4}$); the search-phase dataset configuration
	is detailed in Table~\ref{tab:dataset_setup}.
	
	\begin{table*}[!t]
		\centering
		\caption{20-Run Average Performance Comparison on FER Dataset}
		\label{tab:fer_comparison}
		\renewcommand{\arraystretch}{1.2}
		\begin{small}
			\begin{threeparttable}
				\begin{tabular}{>{\raggedright\arraybackslash}p{2.0cm}
						>{\centering\arraybackslash}p{2.8cm}
						>{\centering\arraybackslash}p{2.8cm}
						>{\centering\arraybackslash}p{2.8cm}
						>{\centering\arraybackslash}p{2.8cm}
						>{\centering\arraybackslash}p{2.8cm}}
					\toprule[1pt]
					Model & Accuracy & Precision & Recall & F1-score & AUC-ROC \\
					\midrule[0.5pt]
					QPSAN
					& $\mathbf{0.7699\pm0.0150}$
					& $\mathbf{0.7822\pm0.0191}$
					& $\mathbf{0.7495\pm0.0410}$
					& $\mathbf{0.7646\pm0.0195}$
					& $\mathbf{0.8181\pm0.0202}$ \\
					MLPSAN-585
					& $0.7390\pm0.0212$
					& $0.7540\pm0.0378$
					& $0.7160\pm0.0488$
					& $0.7324\pm0.0219$
					& $0.7710\pm0.0245$ \\
					MLPSAN-49
					& $0.7359\pm0.0231$
					& $0.7496\pm0.0293$
					& $0.7107\pm0.0477$
					& $0.7285\pm0.0281$
					& $0.7702\pm0.0256$ \\
					ViT
					& $0.7345\pm0.0267$
					& $0.7516\pm0.0312$
					& $0.7025\pm0.0440$
					& $0.7253\pm0.0297$
					& $0.7814\pm0.0319$ \\
					CosViT
					& $0.7200\pm0.0179$
					& $0.7244\pm0.0254$
					& $0.7132\pm0.0408$
					& $0.7177\pm0.0207$
					& $0.7636\pm0.0222$ \\
					LinearViT
					& $0.6874\pm0.0118$
					& $0.7028\pm0.0413$
					& $0.6640\pm0.0641$
					& $0.6788\pm0.0159$
					& $0.7132\pm0.0091$ \\
					QPSAN-Ind
					& $0.6814\pm0.0296$
					& $0.6824\pm0.0450$
					& $0.6890\pm0.0519$
					& $0.6835\pm0.0284$
					& $0.7155\pm0.0278$ \\
					\bottomrule[1pt]
				\end{tabular}
				\begin{tablenotes}
					\footnotesize
					\item Mean $\pm$ std over 20 independent runs across
					seven models on the FER dataset. Bold indicates best
					performance.
					\item This work does not compare against quantum
					attention baselines. Most existing quantum attention
					methods adopt fully quantum architectures and, constrained
					by quantum hardware, have only been validated on
					small-scale benchmarks, rendering them incomparable with
					the setting of this work (see
					Section~\ref{subsec:quantum_attention} for details).
					Furthermore, recent VQC-based quantum attention models
					predominantly target non-vision
					tasks~\cite{tomal2025quantum,chen2025quantuma,smaldone2025hybrid}
					and therefore fall outside the scope of this comparison.
				\end{tablenotes}
			\end{threeparttable}
		\end{small}
	\end{table*}
	
	With optimal configurations, 20 independent runs were conducted;
	detailed results are reported in Table~\ref{tab:fer_comparison}. The
	findings are as follows: (1)~MLPSAN-49 and MLPSAN-585 both perform on
	par with standard ViT, with no significant difference between them;
	(2)~neither CosViT nor LinearViT surpasses the ViT baseline;
	(3)~QPSAN-Ind shows a marked performance drop relative to QPSAN;
	(4)~paired $t$-tests further confirm that QPSAN statistically
	significantly outperforms all comparison models. Detailed test results
	are provided in Table~\ref{tab:pairwise_ttest}.
	
	\begin{table}[!t]
		\centering
		\caption{20-Run Pairwise $t$-Test Results on FER Dataset}
		\label{tab:pairwise_ttest}
		\renewcommand{\arraystretch}{1.2}
		\begin{small}
			\begin{threeparttable}
				\begin{tabular}{>{\raggedright\arraybackslash}p{2.8cm}
						>{\centering\arraybackslash}p{1.2cm}
						>{\centering\arraybackslash}p{1.2cm}
						>{\centering\arraybackslash}p{1.6cm}}
					\toprule[1pt]
					Comparison & Diff & $p$-value & Sig. \\
					\midrule[0.5pt]
					QPSAN vs ViT      & $+3.54\%$ & $0.000017$  & *** \\
					QPSAN vs MLP-49   & $+3.40\%$ & $0.000045$  & *** \\
					QPSAN vs MLP-585  & $+3.09\%$ & $0.000021$  & *** \\
					MLP-49 vs MLP-585 & $-0.31\%$ & $0.687$     & n.s. \\
					MLP-49 vs ViT     & $+0.14\%$ & $0.877$     & n.s. \\
					MLP-585 vs ViT    & $+0.45\%$ & $0.585$     & n.s. \\
					ViT vs CosViT     & $+1.45\%$ & $0.0148$    & * \\
					ViT vs LinearViT  & $+4.71\%$ & $0.0000003$ & *** \\
					QPSAN vs QPSAN-Ind & $+8.85\%$ & $\approx 0$ & *** \\
					\bottomrule[1pt]
				\end{tabular}
				\begin{tablenotes}
					\footnotesize
					\item Diff denotes the difference in validation accuracy
					between the two compared models. $p$-values are based on
					one-tailed paired $t$-tests over 20 independent runs.
					MLP-49 and MLP-585 are abbreviated from MLPSAN-49 and
					MLPSAN-585, denoting ablation models with two-layer and
					three-layer MLP scoring functions, respectively.
					\item Sig.: significance level;\quad n.s.: $p > 0.05$;
					\quad *: $0.01 < p \leq 0.05$;\quad
					**: $0.001 < p \leq 0.01$;\quad ***: $p \leq 0.001$.
				\end{tablenotes}
			\end{threeparttable}
		\end{small}
	\end{table}
	
	The property analysis in Section~\ref{subsec:properties} characterizes
	QPA's structural properties at three levels; the ablation results
	corroborate each level of analysis.
	
	First, Theorems~1--2 and Remark~1 indicate that QPA parameters are
	subject to dual constraints from the linear folding structure of the
	encoding layer and the physical circuit topology, with the effective
	dimension upper-bounded by $d_Q\leq 4$. MLPSAN-49 and MLPSAN-585, by
	contrast, contain 45 and 581 unconstrained free parameters,
	respectively. The universal approximation theorem guarantees that
	sufficiently parameterized MLPs can approximate any continuous
	function; yet, as this experiment demonstrates, increased
	expressiveness does not automatically translate into improved
	generalization---the redundancy introduced by over-parameterization
	may make it harder for the optimization trajectory to converge to
	well-performing optima.
	
	Second, Lemmas~1--2 (Section~\ref{subsec:properties}) characterize the
	structural difference between three-step encoding and single-parameter
	independent encoding from the perspectives of non-separable kernels and
	independent frequency modulation, respectively. Replacing three-step
	encoding with single-parameter independent encoding in the QPSAN-Ind
	ablation yields a performance drop of approximately $8.85\%$ relative
	to QPSAN ($p<0.001$), consistent with the structural differences
	described in Lemmas~1--2.
	
	Taken together, none of the classical baselines across the three hypotheses replicates QPSAN's performance advantage, 
	and the encoding structure ablation
	(QPSAN-Ind) further demonstrates that the three-step encoding makes a
	substantial contribution to performance. This suggests that the
	performance gains may be associated with the overall structural
	inductive bias of the quantum circuit, though the precise mechanism
	remains to be further investigated.
	
	\subsection{Interpretability}
	\label{subsec:visualization_and_interpretability_analysis}
	
	\begin{figure}[!t]
		\centering
		\includegraphics[width=\columnwidth]{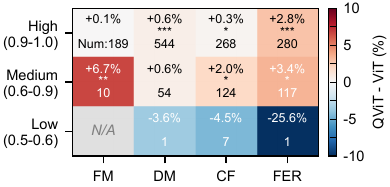}
		\caption{Confidence-stratified accuracy heatmap across datasets.
			Each of the four datasets is partitioned into three confidence
			strata: high \mbox{($0.9$--$1.0$)}, medium ($0.6$--$0.9$), and low
			($0.5$--$0.6$). Stratification and within-stratum accuracy
			computation are performed independently for each run; the
			within-stratum accuracies across 20~runs form paired samples for
			the paired $t$-test. Numbers in the figure denote the average
			number of samples falling into each stratum per run. Cell color
			indicates the validation accuracy difference
			$\text{QPSAN} - \text{ViT}$: deeper red indicates a larger
			accuracy gain of QPSAN over ViT, and deeper blue indicates a
			larger gain of ViT over QPSAN. The percentage in each cell is the
			accuracy difference of QPSAN relative to ViT; the number
			following ``Num'' is the average sample count for that stratum.
			Asterisks denote one-tailed paired $t$-test $p$-value ranges: no
			asterisk indicates a non-significant result ($p>0.05$); $*$
			indicates $0.01 < p \leq 0.05$; $**$ indicates
			$0.001 < p \leq 0.01$; $***$ indicates $p \leq 0.001$
			(e.g., for the high-confidence stratum of \mbox{FER},
			the accuracy gain of QPSAN yields $p = 0.000017$,
			displayed as~$***$).}
		\label{fig:layer_attention}
	\end{figure}
	
	Fig.~\ref{fig:layer_attention} presents the stratified accuracy
	differences ($\text{QPSAN} - \text{ViT}$) to further characterize
	the distribution of performance differences. In the medium-confidence
	stratum, QPSAN outperforms ViT across all four datasets, with
	statistically significant differences on FM ($p = 0.005$), CF
	($p = 0.031$), and FER ($p = 0.029$). In the high-confidence stratum,
	FER shows the most pronounced improvement ($+2.84\%$, $p < 0.001$),
	with DM ($p < 0.001$) and CF ($p = 0.018$) likewise significant. The
	dual significant improvements of FER in both the medium- and
	high-confidence strata are directionally consistent with its overall
	performance difference on the full validation set.
	
	FM shows no significant overall improvement on the full validation
	set; stratified analysis provides one possible explanation. FM
	presents low image classification difficulty, with approximately 95\%
	of validation samples falling in the high-confidence stratum, where
	the accuracy difference between QPSAN and ViT is only $+0.07\%$
	($p = 0.26$). Although FM reaches statistical significance in the
	medium-confidence stratum ($p = 0.005$), this stratum contains on
	average only approximately 10 samples per run, and statistical
	conclusions from this stratum should be interpreted with caution.
	
	\subsection{Further Analysis}
	\label{subsec:further_analysis}
	
	\subsubsection{Noise Robustness}
	\label{subsubsec:noise_robustness_analysis}
	
	\begin{figure}[!t]
		\centering
		\includegraphics[width=\columnwidth]{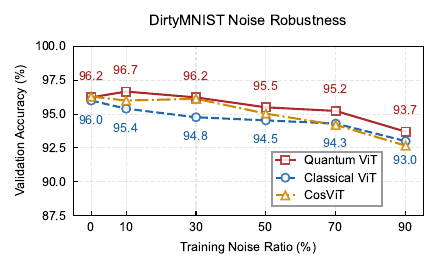}
		\caption{Validation accuracy of \mbox{QPSAN}, \mbox{ViT}, and
			\mbox{CosViT} on the DM dataset under varying proportions of
			training label noise.}
		\label{fig:noise_roubness}
	\end{figure}
	
	\begin{figure}[!t]
		\centering
		\includegraphics[width=\columnwidth]{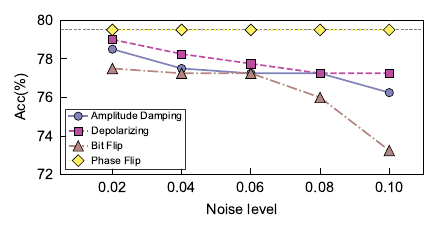}
		\caption{Validation accuracy of \mbox{QPSAN} on the FER dataset
			under four quantum noise channels. Each channel is injected into
			the quantum circuit at varying intensities
			$\gamma_{\text{noise}}$: amplitude damping (\mbox{AD}),
			depolarizing (\mbox{DP}), bit flip (\mbox{BF}), and phase flip
			(\mbox{PF}); the gray dashed line denotes the noise-free
			baseline.}
		\label{fig:quantum_noise}
	\end{figure}
	
	This section evaluates QPSAN's robustness via two categories of noise
	experiments: blur sample noise during training and quantum circuit
	noise during inference.
	
	As the proportion of blurred samples increases
	(Fig.~\ref{fig:noise_roubness}), validation accuracy declines across
	all three models. QPSAN maintains an accuracy advantage over ViT at
	all noise levels ($+0.23\%$ to $+1.47\%$), while CosViT---which also
	employs bounded scoring ($lr=5\times10^{-3}$, $bs=4$ from
	hyperparameter search)---shows overall accuracy degradation
	intermediate between QPSAN and ViT. One possible explanation is that
	QPSAN's bounded scoring property reduces the probability of extreme
	attention scores arising under noise perturbation relative to unbounded
	dot-product attention, thereby partially suppressing 
	the tendency of post-$\operatorname{softmax}$
	weights to concentrate excessively on noisy tokens.
	
	Fig.~\ref{fig:quantum_noise} shows quantum noise injection results
	during inference. As noise intensity increases, 
	all noise types except phase flip (PF) cause accuracy to decline
	to varying degrees. PF noise has negligible
	impact: QPA relies solely on the computational-basis joint measurement
	$\mu=P(|00\rangle)+P(|11\rangle)$, whose value depends only on the
	measurement probability of each basis state and is independent of
	quantum state phase; the PF channel (acting as $Z\otimes Z$) perturbs
	only the phase component without altering any basis-state measurement
	probability, theoretically canceling its effect on the output
	entirely. Following the analysis of Schuld et al., quantum model
	outputs can be expressed as partial Fourier series of the input data,
	with frequency components determined by the encoding gate
	structure~\cite{schuld2021effect}; within this framework, the above
	PF immunity follows strictly from QPA's encoding structure, consistent
	with experimental observations. Bit flip (BF) noise has the largest
	impact, with accuracy dropping 6.25 percentage points at
	$\gamma_{\text{noise}}=0.10$ ($79.50\%\to73.25\%$). Amplitude damping
	(AD) noise has a moderate effect, dropping 3.25 percentage points at
	$\gamma_{\text{noise}}=0.10$. Depolarizing (DP) noise shows saturating
	degradation at $\gamma_{\text{noise}}\geq0.08$. Under
	$\gamma_{\text{noise}}\leq0.10$, accuracy losses from AD and DP remain
	within 3.25\%, while PF causes negligible degradation.
	
	\subsubsection{Complexity Analysis}
	\label{subsubsec:complexity_analysis}
	
	The following complexity analysis covers only the scoring function
	component, across two dimensions: computational complexity and sampling
	complexity.
	
	Classical dot-product attention has time complexity
	$O(H \cdot N^2 \cdot d_h)$. QPA executes $D$ batched circuit calls per
	layer, each processing $H \cdot N^2$ query--key pairs, giving time
	complexity $O(H \cdot N^2 \cdot D)$. Since $D$ is a constant
	independent of input size, both share the same order of asymptotic time
	complexity in sequence length $N$, with identical space complexity
	\mbox{$O(H \cdot N^2)$}.
	
	On a classical simulator, each circuit evaluation operates on a
	$2^{n_q} \times 2^{n_q}$ complex matrix, with the constant-factor
	overhead determined by the ratio of $d_h$ to
	$D \cdot G \cdot 2^{2n_q}$; under our configuration this ratio is
	estimated at approximately $40$--$80\times$, constituting the inherent
	overhead of simulating quantum circuits on classical hardware.
	
	On real quantum hardware, $\mu(q,k) = P(|00\rangle)+P(|11\rangle)$
	must be estimated via repeated measurement. From the binomial variance
	bound $\mathrm{Var}(\hat{\mu}) \leq 1/(4S)$, achieving an estimation
	standard deviation no greater than $\varepsilon$ requires
	$S \geq 1/(4\varepsilon^2)$ shots. Since attention scores are
	normalized by Softmax and are therefore insensitive to absolute
	estimation error, $\varepsilon = 0.05$ suffices for practical accuracy
	under our experimental configuration, corresponding to $S = 100$ shots.
	
	QPA shares the same order of asymptotic computational complexity as
	classical dot-product attention. 
	The overhead described above constitutes the common hardware constraint
	of PQC-based hybrid architectures in the NISQ era, rather than an
	inherent cost of the QPA algorithm itself.
	
	\section{Conclusion}
	\label{sec:conclusion}
	
	This work is the first to employ PQCs as the scoring function of
	multi-head self-attention. With minimal incremental parameters, it
	achieves a structural replacement of the classical scoring function,
	providing a lightweight and theoretically tractable quantum--Transformer
	integration. We systematically establish a mathematical framework for
	the properties of the QPA scoring function, theoretically
	characterizing the structural conditions for representing complex
	scoring patterns, and elucidate the structural inductive bias of the
	quantum circuit over classical alternatives via effective degrees of
	freedom analysis. Experiments show that QPSAN outperforms the baseline
	across multiple vision datasets, with the performance advantage more
	pronounced on more complex datasets. Confidence-stratified analysis
	further reveals that performance gains concentrate on uncertain samples
	near decision boundaries. Ablation studies confirm that classical 
	MLPs with far more parameters
	than the quantum circuit cannot replicate this advantage, suggesting
	that performance gains are associated with the structural inductive
	bias of the quantum circuit rather than parameter scale or nonlinear
	transformation per se.
	
	However, the current work is validated only on binary classification
	tasks and relies on classical simulation; future research should extend
	QPSAN to multi-class tasks to verify the generalizability of these
	findings, and explore deployment on real quantum hardware, providing
	an architectural reference for the transition from NISQ to
	fault-tolerant quantum computing. More broadly, these results preliminarily suggest that
	quantum circuits can serve as structurally controllable,
	performance-verifiable functional modules in neural networks,
	providing empirical support for quantum computing playing a structural
	role in practical AI systems.
	
	\bibliographystyle{IEEEtran}
	\bibliography{cite}

\section*{Proofs of Main Results}
We provide complete proofs of Lemmas~1--2, Properties~1--3, and
Theorems~1--2.

\subsection*{Proof of Lemma~1 (Non-separable kernel)}

\textbf{Step 1: Equivalent rotation angles.}
All three encoding steps act prior to the entanglement layer; each
step applies RY gates independently on each qubit with no cross-qubit
operators, so the rotation angles from each step add linearly on the
same qubit. Summing the three step angles on Qubit~0:
\begin{align}
	&\left(\frac{\pi}{4} + \theta_{\text{s}} q\right)
	+ \gamma_{\text{d}}(q-k) + \gamma_{\text{s}}(q+k) \notag\\
	&\quad= \frac{\pi}{4}
	+ (\theta_{\text{s}}+\gamma_{\text{d}}+\gamma_{\text{s}})q
	+ (\gamma_{\text{s}}-\gamma_{\text{d}})k .
\end{align}
On Qubit~1, the difference encoding applies a negated rotation:
\begin{align}
	&\left(\frac{\pi}{4} + \theta_{\text{s}} k\right)
	- \gamma_{\text{d}}(q-k) + \gamma_{\text{s}}(q+k) \notag\\
	&\quad= \frac{\pi}{4}
	+ (\gamma_{\text{s}}-\gamma_{\text{d}})q
	+ (\theta_{\text{s}}+\gamma_{\text{d}}+\gamma_{\text{s}})k .
\end{align}
Defining $\lambda_1 \triangleq \theta_{\text{s}}+\gamma_{\text{d}}+\gamma_{\text{s}}$
and $\lambda_2 \triangleq \gamma_{\text{s}}-\gamma_{\text{d}}$, the
equivalent rotation angles are
\begin{equation}
	\varphi_0 = \frac{\pi}{4} + \lambda_1 q + \lambda_2 k ,
	\quad
	\varphi_1 = \frac{\pi}{4} + \lambda_2 q + \lambda_1 k .
	\label{seq:equiv-angles}
\end{equation}
The equivalent angles exhibit a cross-symmetric structure: $\lambda_1$
is the self-coefficient (the rotation weight of Qubit~0 for $q$ and of
Qubit~1 for $k$), and $\lambda_2$ is the cross-coefficient (the
rotation weight of Qubit~0 for $k$ and of Qubit~1 for $q$), allowing
each qubit to sense the input feature of the other.

\textbf{Step 2: Derivation of $K_{\text{enc-3}}$.}
Define $\lambda_1' \triangleq \lambda_1/2$ and
$\lambda_2' \triangleq \lambda_2/2$; both will be used throughout
the subsequent proofs. The product state output of the three-step
encoding layer (prior to the entanglement layer) is
\begin{equation}
	|\psi_{\text{enc-3}}(q,k)\rangle
	= \mathrm{RY}(\varphi_0)|0\rangle
	\otimes
	\mathrm{RY}(\varphi_1)|0\rangle .
\end{equation}
For two data points $\bm{x}_1=(q_1,k_1)$ and $\bm{x}_2=(q_2,k_2)$,
the quantum kernel is defined as
$K(\bm{x}_1,\bm{x}_2) = |\langle\psi(\bm{x}_1)|\psi(\bm{x}_2)\rangle|^2$.
The product state has tensor-product structure, so the inner product
factors into a product of per-qubit inner products:
\begin{equation}
	\langle\psi_{\text{enc-3}}(\bm{x}_1)|\psi_{\text{enc-3}}(\bm{x}_2)\rangle
	= \prod_{j=0}^{1}
	\langle 0|\mathrm{RY}^\dagger(\varphi_j(\bm{x}_1))
	\mathrm{RY}(\varphi_j(\bm{x}_2))|0\rangle .
\end{equation}
Using $\mathrm{RY}^\dagger(\alpha)\mathrm{RY}(\beta)=\mathrm{RY}(\beta-\alpha)$,
the $\pi/4$ offset cancels in the difference:
\begin{align}
	\varphi_0(\bm{x}_2)-\varphi_0(\bm{x}_1)
	&= \lambda_1\Delta q+\lambda_2\Delta k , \notag\\
	\varphi_1(\bm{x}_2)-\varphi_1(\bm{x}_1)
	&= \lambda_2\Delta q+\lambda_1\Delta k .
\end{align}
Since $\langle 0|\mathrm{RY}(\theta)|0\rangle=\cos(\theta/2)$, the
inner product is real; squaring the modulus and substituting the
half-angle coefficients yields
\begin{equation}
	K_{\text{enc-3}} =
	\cos^2\!\left(\lambda_1'\Delta q+\lambda_2'\Delta k\right)
	\cdot\cos^2\!\left(\lambda_2'\Delta q+\lambda_1'\Delta k\right) .
	\label{seq:Kenc3-supp}
\end{equation}

\textbf{Step 3: Non-separability.}
We prove that when $\lambda_1\neq 0$ and $\lambda_2\neq 0$,
$K_{\text{enc-3}}$ cannot be decomposed as $f(\Delta q)\cdot g(\Delta k)$.

If $\lambda_1=0$ (i.e., $\lambda_1'=0$), then
$K_{\text{enc-3}}=\cos^2(\lambda_2'\Delta k)\cdot\cos^2(\lambda_2'\Delta q)$
is already separable, so $\lambda_1\neq 0$ is a necessary condition.
Now suppose $\lambda_1\neq 0$ and $\lambda_2\neq 0$, and assume for
contradiction that $K_{\text{enc-3}}$ is separable. Since
$K_{\text{enc-3}}(0,0)=1$, by continuity there exists a neighborhood
of $(0,0)$ in which $K_{\text{enc-3}}>0$. Taking the logarithm in
this neighborhood, separability requires
\begin{equation}
	\ln K_{\text{enc-3}}(\Delta q,\Delta k)
	= \ln f(\Delta q) + \ln g(\Delta k) ,
\end{equation}
so the mixed partial derivative must be identically
zero~\cite{scholkopf2002learning}. Direct computation, however, gives
\begin{multline}
	\frac{\partial^2\ln K_{\text{enc-3}}}
	{\partial(\Delta q)\partial(\Delta k)}
	= -2\lambda_1'\lambda_2'
	\left[\sec^2\!\left(
	\lambda_1'\Delta q+\lambda_2'\Delta k\right)\right.\\
	\left.
	+\sec^2\!\left(
	\lambda_2'\Delta q+\lambda_1'\Delta k\right)\right] .
\end{multline}
Since $\sec^2(\cdot)\geq 1$, the bracketed term is always $\geq 2$.
Whenever $\lambda_1\neq 0$ and $\lambda_2\neq 0$, this mixed partial
is strictly nonzero, contradicting separability.
\hfill$\blacksquare$

\subsection*{Proof of Property~1 (Boundedness)}
By definition, $\mu(q,k) = P(|00\rangle)+P(|11\rangle)$, where
$P(|ab\rangle)\geq 0$ and $\sum_{a,b}P(|ab\rangle)=1$. Therefore
\begin{equation}
	0 \leq P(|00\rangle)+P(|11\rangle)
	\leq \textstyle\sum_{a,b}P(|ab\rangle) = 1 .
\end{equation}
\hfill$\blacksquare$

\subsection*{Proof of Property~2 (Asymmetry of the full circuit,
	$\alpha\neq 0$)}
Define the two-qubit swap operator $\mathrm{SWAP}$. If the full circuit
satisfies $\mathrm{SWAP}\cdot U(q,k)\cdot\mathrm{SWAP}=U(k,q)$, then
since $\mathrm{SWAP}|00\rangle=|00\rangle$ and the measurement operator
$\Pi=|00\rangle\langle 00|+|11\rangle\langle 11|$ satisfies
$\mathrm{SWAP}\cdot\Pi\cdot\mathrm{SWAP}=\Pi$, the full trace computation
gives:
\begin{align}
	\mu(q,k)
	&= \operatorname{Tr}[\Pi\cdot U(q,k)|00\rangle\langle 00|
	U^\dagger(q,k)] \notag\\
	&= \operatorname{Tr}[\Pi\cdot U(q,k)\cdot\mathrm{SWAP}
	|00\rangle\langle 00|\mathrm{SWAP}\cdot U^\dagger(q,k)] \notag\\
	&= \operatorname{Tr}[\Pi\cdot\mathrm{SWAP}\cdot U(k,q)
	|00\rangle\langle 00|U^\dagger(k,q)\cdot\mathrm{SWAP}] \notag\\
	&= \operatorname{Tr}[\mathrm{SWAP}\cdot\Pi\cdot\mathrm{SWAP}
	\cdot U(k,q)|00\rangle\langle 00|U^\dagger(k,q)] \notag\\
	&= \mu(k,q) .
\end{align}
SWAP covariance is therefore a sufficient condition for symmetry.

The encoding layer
$U_{\text{enc}}=\mathrm{RY}(\varphi_0)\otimes\mathrm{RY}(\varphi_1)$
is SWAP covariant: by~\eqref{seq:equiv-angles},
$\varphi_0(k,q)=\varphi_1(q,k)$, so swapping $q$ and $k$ is
equivalent to swapping the RY gates on the two qubits, giving
$\mathrm{SWAP}\cdot U_{\text{enc}}(q,k)\cdot\mathrm{SWAP}
=U_{\text{enc}}(k,q)$. The Mixer layer
$\mathrm{RX}(2\beta)\otimes\mathrm{RX}(2\beta)$ applies the same
operation to both qubits and trivially satisfies
$\mathrm{SWAP}\cdot U_{\text{mix}}\cdot\mathrm{SWAP}=U_{\text{mix}}$.
However, the entanglement layer
\begin{equation}
	\scalebox{0.92}{$U_{\text{ent}}=\mathrm{CNOT}_{0\to1}
		\cdot(I\otimes\mathrm{RY}(\alpha(q+k)))
		\cdot\mathrm{CNOT}_{1\to0}$}
	\label{seq:uent}
\end{equation}
breaks SWAP symmetry. Expanding the conjugation gate by gate using
$\mathrm{SWAP}^2=I$ (applying successively
$\mathrm{SWAP}\cdot\mathrm{CNOT}_{i\to j}\cdot\mathrm{SWAP}
=\mathrm{CNOT}_{j\to i}$ and
$\mathrm{SWAP}\cdot(A\otimes B)\cdot\mathrm{SWAP}=B\otimes A$), we
obtain:
\begin{align}
	&\mathrm{SWAP}\cdot U_{\text{ent}}\cdot\mathrm{SWAP} \notag\\
	&=\mathrm{CNOT}_{1\to0}
	\cdot(\mathrm{RY}(\alpha(q+k))\otimes I)
	\cdot\mathrm{CNOT}_{0\to1} .
	\label{seq:uent-swap}
\end{align}
Compared with~\eqref{seq:uent}, the conjugated version has the RY
gate acting on the first qubit while the original acts on the second;
these are unequal as matrices when $\alpha(q+k)\neq 0$. Since
$U_{\text{ent}}$ depends on the input only through $\alpha(q+k)$ and
$q+k=k+q$, we have $U_{\text{ent}}(k,q)=U_{\text{ent}}(q,k)$, and
hence
$\mathrm{SWAP}\cdot U_{\text{ent}}(q,k)\cdot\mathrm{SWAP}
\neq U_{\text{ent}}(k,q)$.

Expanding the SWAP conjugate of the full circuit layer by layer, and
substituting the Mixer layer's invariance and the encoding layer's
covariance:
\begin{align}
	&\mathrm{SWAP}\cdot U(q,k)\cdot\mathrm{SWAP} \notag\\
	&=U_{\text{mix}}\cdot
	[\mathrm{SWAP}\cdot U_{\text{ent}}(q,k)\cdot\mathrm{SWAP}]
	\cdot U_{\text{enc}}(k,q) .
\end{align}
If the full circuit were SWAP covariant, this would equal
$U_{\text{mix}}\cdot U_{\text{ent}}(k,q)\cdot U_{\text{enc}}(k,q)$,
contradicting~\eqref{seq:uent-swap}; hence the full circuit
structurally breaks SWAP covariance.

\textbf{From structural asymmetry to measurement asymmetry.}
Define $\Delta(q,k)=\mu(q,k)-\mu(k,q)$. By the algebraic structure
of PQC rotation gates, $\mu(q,k)$ is a multivariate trigonometric
polynomial~\cite{schuld2021effect}, and so is $\Delta(q,k)$. Taking
the partial derivative with respect to $\alpha$ and evaluating at
$\alpha=0$:
\begin{equation}
	\frac{\partial\Delta}{\partial\alpha}\bigg|_{\alpha=0}
	=\frac{\partial\mu(q,k)}{\partial\alpha}\bigg|_{\alpha=0}
	-\frac{\partial\mu(k,q)}{\partial\alpha}\bigg|_{\alpha=0} .
\end{equation}
Since $\mathrm{RY}(\alpha(q+k))$'s dependence on $\alpha$ enters only
through the factor $q+k$, and $q+k=k+q$, the difference of the two
partial derivatives does not originate from this factor; it arises
entirely from the exchange of encoding angles.
By~\eqref{seq:equiv-angles}, $\varphi_0(q,k)=\varphi_1(k,q)$, so
$q\leftrightarrow k$ is equivalent to swapping the encoding angles on
the two qubits. Since
\mbox{$U_{\text{ent}}(\alpha=0)=\mathrm{CNOT}_{0\to1}\cdot\mathrm{CNOT}_{1\to0}$} acts asymmetrically on the two qubits (it is not equal to SWAP), when
$\lambda_1\neq\lambda_2$ the exchange of encoding angles causes
\mbox{$\partial\mu(q,k)/\partial\alpha|_{\alpha=0}\neq\partial\mu(k,q)/\partial\alpha|_{\alpha=0}$},
so $\partial\Delta/\partial\alpha|_{\alpha=0}\not\equiv 0$, and
$\Delta$ as an analytic function of the parameters is not identically
zero. Since $\Delta\equiv 0$ is equivalent to the parameters
$(\theta_{\text{s}},\gamma_{\text{d}},\gamma_{\text{s}},\alpha,\beta)$
satisfying a finite set of analytic equations, and the zero set of a
nontrivial analytic function has measure
zero~\cite{krantz2002primer}, this solution set has measure zero in
the parameter space $\mathbb{R}^5$.\hfill$\blacksquare$

\subsection*{Proof of Property~3 (Non-monotonicity of the full
	circuit over $\mathbb{R}^2$)}
$|q-k|$ measures the distance between query and key; in classical
attention mechanisms, scores typically decrease monotonically with
distance. We prove that QPA breaks this assumption.

\textbf{Part I (encoding sub-circuit).}
By Lemma~2, fixing $k$ and setting $\delta=q-k$ so that
$q+k=\delta+2k$, substitution gives
\begin{equation}
	\mu = \frac{1}{2}+\frac{1}{4}\cos(\omega_{\text{d}}\delta)
	-\frac{1}{4}\sin\!\big(\omega_{\text{s}}(\delta+2k)\big) .
\end{equation}
When $\omega_{\text{d}}\neq 0$ or $\omega_{\text{s}}\neq 0$, this is
a persistently oscillating function of $\delta$ that does not converge
as $\delta\to+\infty$. By Property~1, $\mu\in[0,1]$ is bounded; a
monotone bounded function must have a finite limit---a contradiction.
Hence $\mu$ is non-monotone with respect to $\delta=q-k$.
In particular, restricted to $\delta>0$, both
$\cos(\omega_{\text{d}}\delta)$ and
$\sin(\omega_{\text{s}}(\delta+2k))$ continue to oscillate
persistently, so $\mu$ is non-monotone on $\delta>0$ as well. If
$\mu$ were monotone with respect to $|q-k|$, it would have to be
monotone on $\delta>0$---a contradiction. Hence the encoding
sub-circuit's $\mu$ is non-monotone with respect to $|q-k|$.

\textbf{Part II (full circuit).}
The full QPA circuit consists of rotation gates (RY, RX) and CNOTs.
The inputs $q$ and $k$ enter linearly only through rotation gates:
the encoding layer introduces rotation angles
$\varphi_0=\pi/4+\lambda_1 q+\lambda_2 k$ and
$\varphi_1=\pi/4+\lambda_2 q+\lambda_1 k$
(from~\eqref{seq:equiv-angles}), the entanglement layer introduces
rotation angle $\alpha(q+k)$, and the Mixer layer
$\mathrm{RX}(2\beta)$ is independent of the input. The matrix entries
of RY gates are $\cos(\cdot/2)$ and $\sin(\cdot/2)$, and CNOT is a
fixed permutation matrix; therefore the measurement expectation
$\mu(q,k)=\langle\psi|\Pi|\psi\rangle$ can be expanded as a
multivariate trigonometric polynomial~\cite{schuld2021effect}:
\begin{align}
	\mu(q,k) = \sum_{n,m}\!\Big[
	&a_{n,m}\cos\!\big(n\Omega_1 q + m\Omega_2 k\big) \notag\\
	+&b_{n,m}\sin\!\big(n\Omega_1 q + m\Omega_2 k\big)
	\Big] ,
\end{align}
where the sum is finite and the coefficients $a_{n,m},b_{n,m}$ are
functions of the circuit parameters.

Setting $k=0$, $\mu(q,0)$ reduces to a univariate trigonometric
polynomial in $q$. Except on a degenerate parameter set of measure
zero, this polynomial contains at least one nonzero high-frequency
term and oscillates persistently on $\mathbb{R}$ without converging.
By Property~1, $\mu\in[0,1]$ is bounded; a monotone bounded function
must have a finite limit---a contradiction. Hence $\mu(q,0)$ is
non-monotone with respect to $|q|$, i.e., the full circuit's $\mu$
is non-monotone with respect to $|q-k|$.

\textbf{Remark (bounded input domain).}
The above result applies to the global input domain $\mathbb{R}^2$.
If inputs are normalized to a bounded interval whose length is smaller
than the minimal positive period of $\mu(q,0)$, then $\mu$ may be
locally monotone on that interval. In practice, inputs are typically
constrained by LayerNorm; the global non-monotonicity manifests
as the model's nonlinear response capability across different input
scales, rather than a guarantee for every local interval.
\hfill$\blacksquare$


\subsection*{Proof of Theorem~1 (Effective Degrees of Freedom of the
	Encoding Layer)}

By Lemma~2, $\mu(q,k)$ is entirely determined by two frequencies
$(\omega_{\text{d}},\omega_{\text{s}})$, where
$\omega_{\text{d}}=\theta_{\text{s}}+2\gamma_{\text{d}}$ and
$\omega_{\text{s}}=\theta_{\text{s}}+2\gamma_{\text{s}}$.
The Jacobian matrix of $(\omega_{\text{d}},\omega_{\text{s}})^\top$
with respect to $(\theta_{\text{s}},\gamma_{\text{d}},\gamma_{\text{s}})^\top$
is
\begin{equation}
	\bm{J} = \begin{pmatrix} 1 & 2 & 0 \\ 1 & 0 & 2 \end{pmatrix} .
\end{equation}
The determinant of the leading $2\times 2$ submatrix is $-2\neq 0$,
so $\operatorname{rank}(\bm{J})=2$; the mapping from 3 encoding
parameters to 2 frequencies is surjective but not injective. Since
$\cos(\omega_{\text{d}}(q-k))$ is an even function of
$\omega_{\text{d}}$, restricting the parameter domain to
$\omega_{\text{d}}\geq 0$ does not change the image of the function
class.

Under this restriction, set $u=q-k$ and $v=q+k$ (the Jacobian
determinant is $2\neq 0$, so $u$ and $v$ can vary independently). If
two parameter sets give the same function:
\begin{align}
	& \cos(\omega_{\text{d}}^{(1)}u)-\sin(\omega_{\text{s}}^{(1)}v) \notag\\
	&\equiv \cos(\omega_{\text{d}}^{(2)}u)-\sin(\omega_{\text{s}}^{(2)}v)
	\quad \forall\, u,v ,
\end{align}
fixing $v$ and differentiating both sides twice at $u=0$ gives
$(\omega_{\text{d}}^{(1)})^2=(\omega_{\text{d}}^{(2)})^2$; combined
with $\omega_{\text{d}}\geq 0$, this yields
$\omega_{\text{d}}^{(1)}=\omega_{\text{d}}^{(2)}$. Fixing $u$ and
differentiating both sides at $v=0$ gives
$\omega_{\text{s}}^{(1)}=\omega_{\text{s}}^{(2)}$. The mapping thus
constitutes an injective embedding on $\omega_{\text{d}}\geq 0$, and
the dimension of the function class $\mathcal{F}_Q$ is~2.
\hfill$\blacksquare$

\subsection*{Proof of Theorem~2 (Bounds on the Effective Degrees of
	Freedom of the Full Circuit)}

\textbf{Upper bound $d_Q\leq 4$.}
Define the reparametrization map
\begin{equation}
	\varphi:\,\bm{\Theta}=(\theta_{\text{s}},\gamma_{\text{d}},\gamma_{\text{s}},\alpha,\beta)
	\;\longmapsto\;(\omega_{\text{d}},\omega_{\text{s}},\alpha,\beta) ,
\end{equation}
with Jacobian matrix
\begin{equation}
	\bm{J}_\varphi =
	\begin{pmatrix}
		1 & 2 & 0 & 0 & 0 \\
		1 & 0 & 2 & 0 & 0 \\
		0 & 0 & 0 & 1 & 0 \\
		0 & 0 & 0 & 0 & 1
	\end{pmatrix} .
\end{equation}
The first two rows have rank~2, and the last two rows are independent
identity components, so $\operatorname{rank}(\bm{J}_\varphi)=4$.
By the chain rule and the matrix rank
inequality~\cite{horn2012matrix},
\begin{equation}
	d_Q = \max_{\bm{\Theta}}\operatorname{rank}(\bm{J}_{\text{full}}(\bm{\Theta}))
	\leq \operatorname{rank}(\bm{J}_\varphi) = 4 .
\end{equation}

\textbf{Lower bound $d_Q\geq 2$.}
Set $\alpha=0$ and \mbox{$\beta=0$}. The entanglement layer degenerates to
a fixed unitary \mbox{$U_V=\mathrm{CNOT}_{0\to1}\cdot\mathrm{CNOT}_{1\to0}$};
direct computation gives
\begin{equation}
	\Pi_{U_V} = U_V^\dagger\Pi U_V
	= |00\rangle\langle00|+|10\rangle\langle10|
	= I\otimes|0\rangle\langle 0| ,
\end{equation}
so the measurement degenerates to a single-qubit projection onto
Qubit~1. By~\eqref{seq:equiv-angles}, the state of Qubit~1 is
$\mathrm{RY}(\varphi_1)|0\rangle$ with
$\varphi_1=\pi/4+\lambda_2 q+\lambda_1 k$, giving
\begin{equation}
	\mu = \cos^2\!\left(\frac{\pi}{8}
	+\lambda_2' q+\lambda_1' k\right) .
\end{equation}
The partial derivative with respect to $\lambda_1'$ has $k$ as a
factor, and that with respect to $\lambda_2'$ has $q$ as a factor;
for general $(q,k)$ with $q\neq 0$ and $k\neq 0$, these are linearly
independent, so the effective dimension is~2, giving
$d_Q\geq 2$.\hfill$\blacksquare$

\section*{Complete Experimental Results}

Table~\ref{tab:supp_full_results} reports the complete evaluation
metrics of QPSAN and the classical ViT baseline over 20 independent
runs on all four datasets. The main paper presents these results
in figure form; the full numerical values are provided here.

\begin{table*}[!t]
	\centering
	\caption{Full 20-Run Average Performance Comparison Across
		All Four Datasets}
	\label{tab:supp_full_results}
	\renewcommand{\arraystretch}{1.2}
	\begin{small}
		\begin{threeparttable}
			\begin{tabular}{>{\raggedright\arraybackslash}p{1.6cm}
					>{\centering\arraybackslash}p{3.0cm}
					>{\centering\arraybackslash}p{3.0cm}
					>{\centering\arraybackslash}p{3.0cm}
					>{\centering\arraybackslash}p{3.0cm}
					>{\centering\arraybackslash}p{3.0cm}}
				\toprule[1pt]
				Dataset & Accuracy & Precision & Recall & F1-score & AUC-ROC \\
				\midrule[0.5pt]
				C-FM
				& $0.9793\pm0.0054$
				& $0.9868\pm0.0056$
				& $0.9715\pm0.0096$
				& $0.9791\pm0.0053$
				& $\boldsymbol{0.9915\pm0.0030}$ \\
				Q-FM
				& $\boldsymbol{0.9822\pm0.0087}$
				& $\boldsymbol{0.9869\pm0.0079}$
				& $\boldsymbol{0.9775\pm0.0122}$
				& $\boldsymbol{0.9821\pm0.0086}$
				& $0.9899\pm0.0032$ \\
				\midrule[0.3pt]
				C-DM
				& $0.9187\pm0.0091$
				& $0.9139\pm0.0115$
				& $\boldsymbol{0.9247\pm0.0133}$
				& $0.9191\pm0.0088$
				& $0.9640\pm0.0049$ \\
				Q-DM
				& $\boldsymbol{0.9239\pm0.0074}$
				& $\boldsymbol{0.9250\pm0.0111}$
				& $0.9228\pm0.0123$
				& $\boldsymbol{0.9238\pm0.0073}$
				& $\boldsymbol{0.9703\pm0.0035}$ \\
				\midrule[0.3pt]
				C-CF
				& $0.7806\pm0.0144$
				& $0.8598\pm0.0322$
				& $\boldsymbol{0.6743\pm0.0557}$
				& $0.7534\pm0.0253$
				& $0.8865\pm0.0092$ \\
				Q-CF
				& $\boldsymbol{0.7881\pm0.0183}$
				& $\boldsymbol{0.8880\pm0.0392}$
				& $0.6638\pm0.0620$
				& $\boldsymbol{0.7566\pm0.0301}$
				& $\boldsymbol{0.9018\pm0.0139}$ \\
				\midrule[0.3pt]
				C-FER
				& $0.7345\pm0.0260$
				& $0.7516\pm0.0312$
				& $0.7025\pm0.0440$
				& $0.7253\pm0.0297$
				& $0.7814\pm0.0319$ \\
				Q-FER
				& $\boldsymbol{0.7699\pm0.0150}$
				& $\boldsymbol{0.7822\pm0.0191}$
				& $\boldsymbol{0.7495\pm0.0410}$
				& $\boldsymbol{0.7646\pm0.0195}$
				& $\boldsymbol{0.8181\pm0.0202}$ \\
				\bottomrule[1pt]
			\end{tabular}
			\begin{tablenotes}
				\footnotesize
				\item C denotes the classical ViT baseline; Q denotes
				the quantum QPSAN model. FM, DM, CF, and FER correspond
				to FashionMNIST, DirtyMNIST, CIFAR-10, and FER2013,
				respectively. All metrics are reported as mean $\pm$
				standard deviation over 20 independent runs. Bold
				indicates the better result within each dataset pair.
			\end{tablenotes}
		\end{threeparttable}
	\end{small}
\end{table*}

\end{document}